\documentclass[useAMS,usenatbib]{mn2e}
\usepackage{graphicx}
\usepackage{txfonts}
\usepackage{natbib}
\usepackage{color}

\newcommand{\xte}{{\it RXTE}}

\newcommand{\xmm}{{\it XMM-Newton}}

\def\gsim{\mathrel{\hbox{\rlap{\hbox{\lower4pt\hbox{$\sim$}}}\hbox{$>$}}}}
\def\lsim{\mathrel{\hbox{\rlap{\hbox{\lower4pt\hbox{$\sim$}}}\hbox{$<$}}}}

\def\xmm{{\it XMM-Newton}}
\def\E{\mathrm{E}}
\def\V{\mathrm{V}}
\def\gsim{\mathrel{\hbox{\rlap{\hbox{\lower4pt\hbox{$\sim$}}}\hbox{$>$}}}}
\def\lsim{\mathrel{\hbox{\rlap{\hbox{\lower4pt\hbox{$\sim$}}}\hbox{$<$}}}}
\begin{document}
   \title[]{The Ubiquity of the rms-flux relation in Black Hole X-ray Binaries}

   \author[]{L. M. Heil$^{1}$, S. Vaughan$^{1}$, P. Uttley$^{2}$ \\
   $^{1}$ X-Ray and Observational Astronomy Group, University of
   Leicester,  Leicester LE1 7RH\\
	$^{2}$ Astronomical Institute ``Anton Pannekoek", University of Amsterdam, Postbus 94249, 1090 GE Amsterdam, The Netherlands}

   \date{Draft \today}

   \pagerange{\pageref{firstpage}--\pageref{lastpage}} \pubyear{2002}

   \maketitle
   
   \label{firstpage}

   \begin{abstract}
	We have investigated the short term linear relation between the rms variability and the flux in 1,961 observations of 9 black hole X-ray binaries. The rms-flux relation for the 1-10 Hz range is ubiquitously observed in any observation with good variability signal to noise ($>$3$\%$ 1-10 Hz fractional rms). This concurs with results from a previous study of Cygnus X-1 \citep{Gleissner04}, and extends detection of the rms-flux relation to a wider range of states. We find a strong dependence of the flux intercept of the rms-flux relation on source state; as the source transitions from the hard state into the hard intermediate state the intercept becomes strongly positive. We find little evidence for flux dependence of the broad-band noise within the PSD shape, excepting a small subset of observations from one object in an anomalous soft-state. We speculate that the ubiquitous linear rms-flux relation in the broad band noise of this sample, representing a range of different states and objects, indicates that its formation mechanism is an essential property of the luminous accretion flow around black holes.
   \end{abstract}

 

\section{Introduction}
\label{sect:intrormsbh}

	Accreting compact objects such as black hole X-ray binary systems (BHBs) are known to display strong short-term variability, which is dependent on the accretion state. Systematic changes in the power spectra are observed as a source moves through different states over the course of an outburst \citep{Belloni10}. The similarities between these changes for different sources suggest a common production mechanism, so investigation and comparison of the short-term timing properties of different BHB systems is an important tool in understanding accretion in these sources.

	One property which appears to be common to many accreting compact objects is a linear relationship between the rms variability of a source and the count rate. Initially discovered in observations of one black hole and one neutron star X-ray binary \citep[Cyg X-1 and SAX 1808.4-3658 respectively;][]{Uttley01}, this rms-flux relation has since been studied in a few other observations of BHBs, AGN, a neutron star X-ray binary and a ULX \citep[see e.g.][]{Uttley01, Uttley04, Gaskell04, Uttley05, Heil10, Heil11} and studied in detail for all Rossi X-ray Timing Explorer (\xte) observations of Cyg X-1 up to early 2003 \citep{Gleissner04}. Optical observations of three BHBs in the low-hard state have also shown the rms-flux relation \citep{Gandhi09}. The range of compact objects from which this relationship has now been observed is striking and suggests that it could be a ubiquitous property of luminous accretion flows. Before this can be established it needs to be observed in a much wider range of sources and observations. 

	The rms-flux relation in Cygnus X-1 has been detected over timescales from seconds to years \citep{Gleissner04}. This consistency and range of timescales is difficult to explain using models employing additions of independent shots; a connection is required between variability occurring over timescales of years and that happening in seconds. A propagating fluctuations model, such as that suggested by \cite{Lyubarskii97}, where long-term variations from the outer reaches of the disc propagate inwards, coupling to short-term variations at smaller annuli, provides a natural explanation for the observed correlation \citep{Uttley01}. This modulation implies that the flux distribution from the source should be log-normal \citep{Uttley05}. Log-normal flux distributions have been observed from various BHBs and AGN \citep[see e.g.][]{Uttley05, Gaskell04, Giebels09}.

	Although the short-term relation has been studied in many observations of Cyg X-1 and a selected few other BHBs, it has not been consistently observed in the full range of states exhibited by transient sources. The rms-flux relation may be ubiquitously present in all states, or alternatively the relation may not be seen in a particular state suggesting that there is something essentially different in the nature of the accretion flow in between the states. In order to test the ubiquity of the rms-flux relation in BHBs we have utilised the full range of data from 9 BHBs in the \xte~ archive.

\begin{table*}
\label{tab:sources}
\centering
\begin{tabular}{lrrrl}
\hline\hline

Target         &  Total obs.  & Obs. $>$3$\%$ $\sigma_{frac}$ & Good obs. & Obs. IDs.\\
(1)            & (2)   & (3)   & (4)  & (5)\\

\hline

GX 339-4       & 487   & 376   &  235 &  P20056, P20181, P30165, P40031, P40105, P40108, P50117, P70109, P90704, P92035,\\
  & & & & P92052, P92085, P92428, P92704, P93409, P93702, P94331, P95409 (until 27th May 2010)\\
XTE J1118+480  & 48    & 30    &  26  &  P50133, P50407 \\ 
GS 1354-64     & 8     & 7     &  7   &  P20431, P30401 \\
4U 1543-475    & 97    & 25    &  12  &  P70124, P70128, P70133 \\
XTE J1550-564  & 153   & 106   &  81  &  P40401, P50134, P50135, P50137, P60428, P80135 \\
XTE J1650-500  & 177   & 69    &  39  &  P60113, P60118, P70124 \\
GRO J1655-40   & 462   & 140   &  93  &  P10261, P20402, P90058, P90428, P91404, P91702, P91704 \\
H1743-322      & 406   & 297   &  129 &  P80135, P80137, P80144, P80146, P90058, P90115, P90421, P91050, P91428, P92047, \\
 & & & & P93427, P94431, P95405 \\
XTE J1859+226  & 123   & 55    &  12  & P40122, P40124, P40440, P50401 \\
\hline
Total          & 1961  & 1105  &  634 & \\
\hline
\hline
\end{tabular}
\caption{Note: The 1998 outburst of XTE J1550-564 is excluded from this data set due to the strong QPO; The 2004 outburst of GX 339-4 and 2005 outburst of XTE J1118+480 are mostly excluded as there are very few observations in the modes processed for this work.}
\end{table*}

\section{Data Analysis}
\label{sect:datarmsbh}

\subsection{Lightcurve extraction}

We have utilised data from the \xte~ archive for the 9 BHBs given in Table \ref{tab:sources} taken by the Proportional Counter Array (PCA) until May 2010. The analysis was limited to those observations taken in either Binned, Single Bit or Event modes; observations with no data in any of these modes have been excluded from the data set. If observations were made with some or all of these modes in the energy range required, then Binned or Event mode data was used in preference to Single Bit mode to allow finer control over the energy range selection. We have also excluded observations without a file in Standard 2 mode, as these data were used for the background subtraction (this mainly applies to a few observations taken when there were problems with the hardware on \xte). Proposal Identifiers with fewer than six observations are also excluded from the data set. These objects have been selected as they are all well known black hole X-ray binaries, which are bright in outburst and have a number of observations within the \xte~ archive. This is not a complete sample of low mass BHBs in the \xte~ archive, but is chosen to be representative of these sources in general. All lightcurves have been extracted and then binned to around 3 ms (depending slightly on the time resolution of the data mode).  In order to adjust for the differing combinations of detectors used in observations taken with the PCA all results presented here are normalised to one PCU (proportional counter unit). All of these observations are taken when the object is in outburst; in quiescence the count rates are generally too low for detailed timing analysis. Consequently the observations are not evenly sampled in time but do represent a range of states for each source. Due to problems caused by a strong quasi-periodic oscillation (QPO - discussed in Section \ref{sec:statedep}) most observations from the 1998 outburst of XTE J1550-564 are excluded from the sample \citep{Heil11}.

In general the analysis was performed on data extracted from the energy range closest to 2-13 keV, but this differs slightly between observations due to the different data modes used and variations in detector operation. As Standard 2 channel 0 is thought to introduce artificial variability into the lightcurves \citep[see e.g.][]{Gleissner04} this channel was excluded. The lightcurves were then background corrected using simulated background lightcurves produced by the ftool PCABACKEST. We applied the bright source model to all observations, despite some having count rates below the 40 ct/s/PCU threshold where the ``faint source" model is usually preferred. Our choice ensures some consistency in the analysis of observations on either side of the threshold, but experiments using both background models confirm that the choice made little difference to the overall results Observations with a count rate $<$ 10 ct/s/PCU are excluded from the sample.

\begin{figure*}
\begin{center}
$
\begin{array}{lr}
	\includegraphics[width=5.5cm, angle=90]{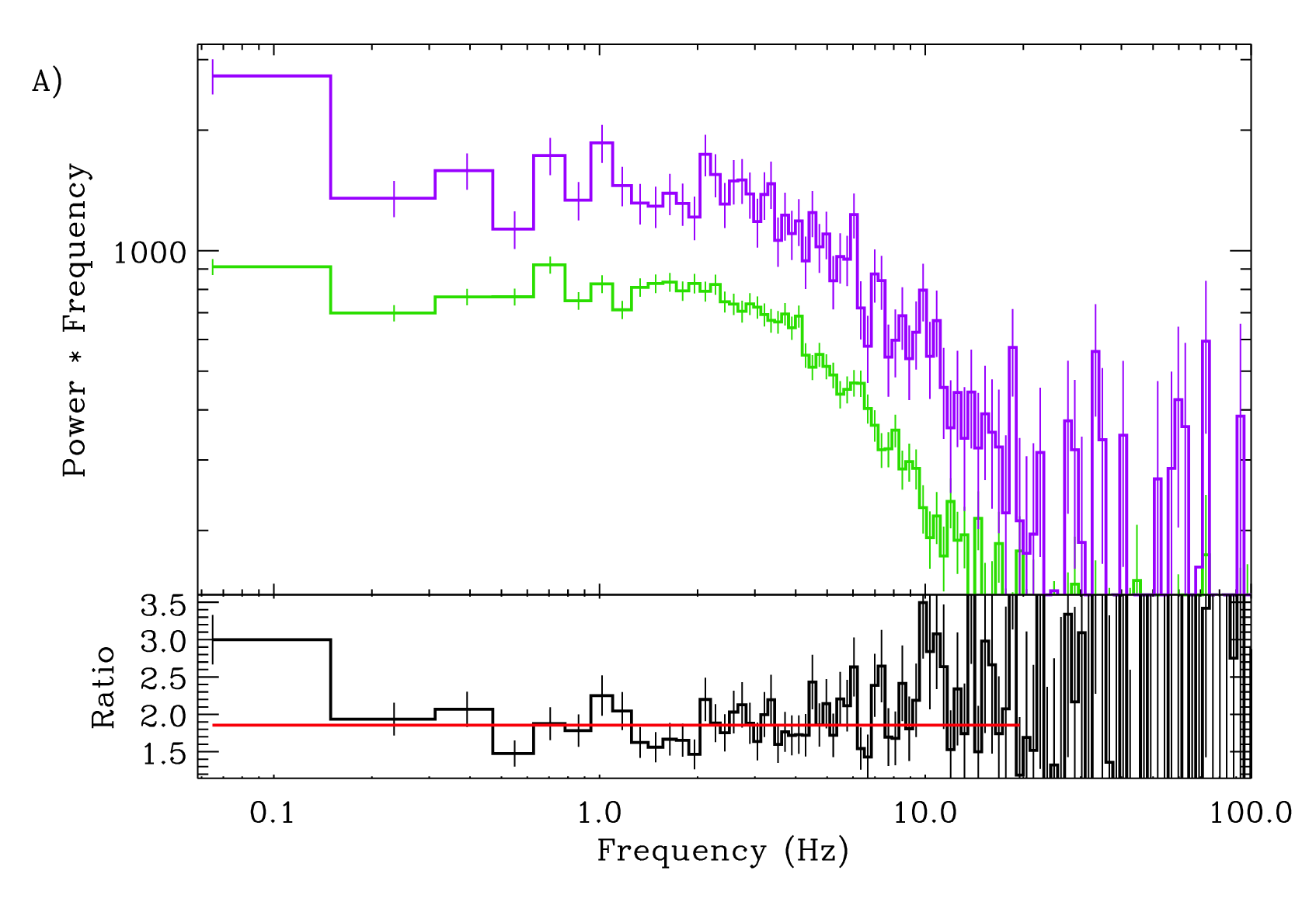} &
	\includegraphics[width=5.5cm, angle=90]{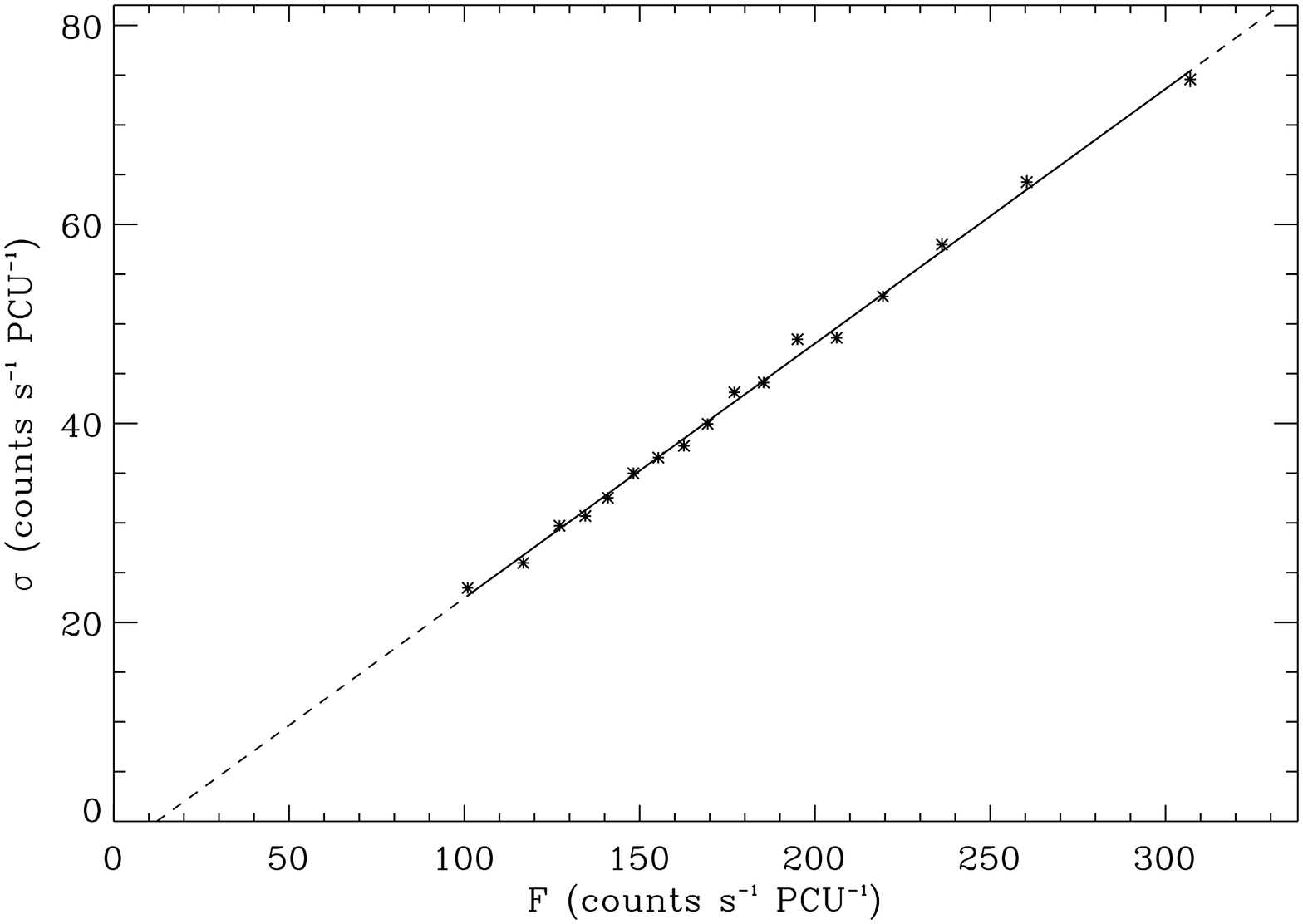}
\end{array}$
$
\begin{array}{lr}
	\includegraphics[width=5.5cm, angle=90]{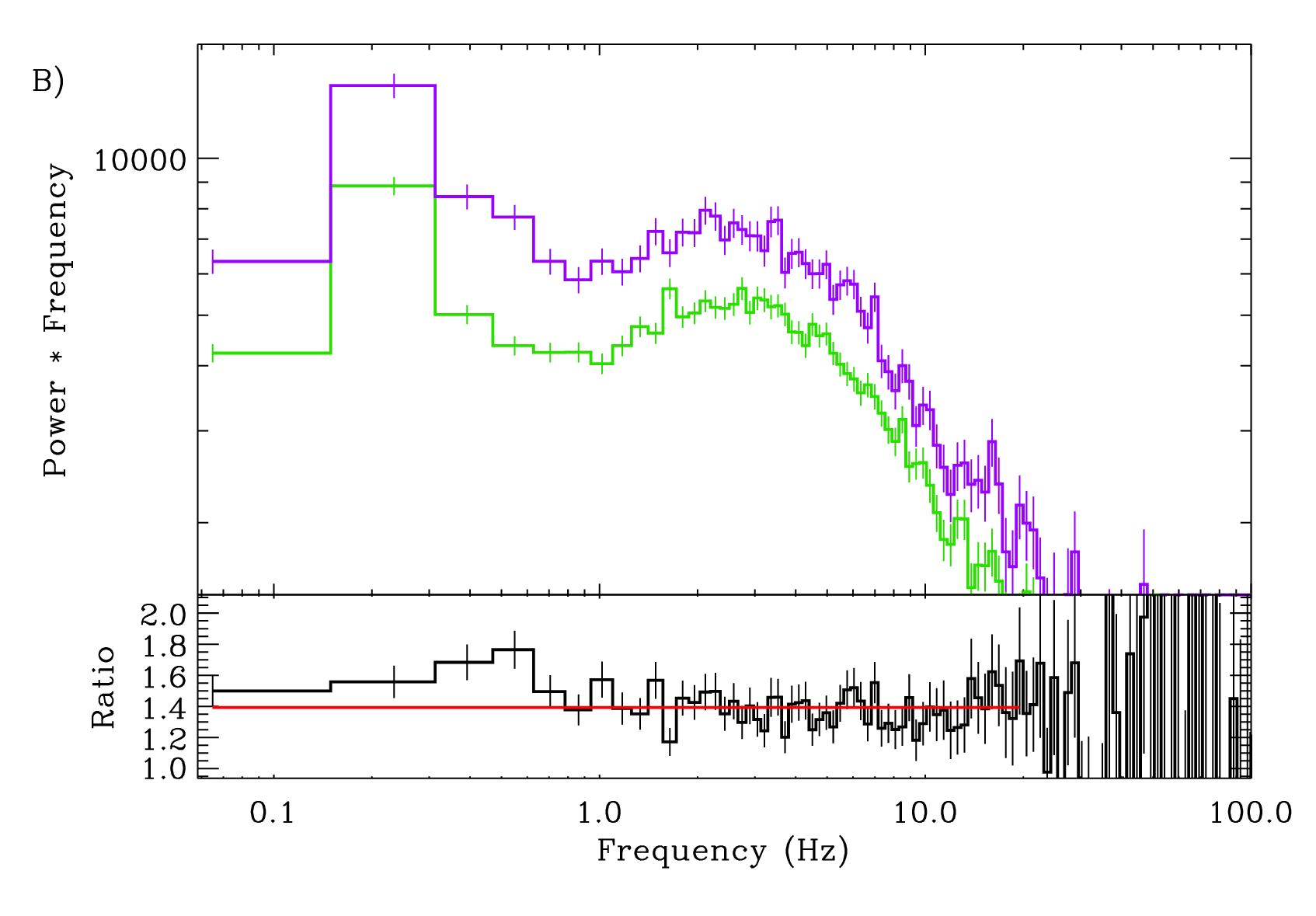} &
	\includegraphics[width=5.5cm, angle=90]{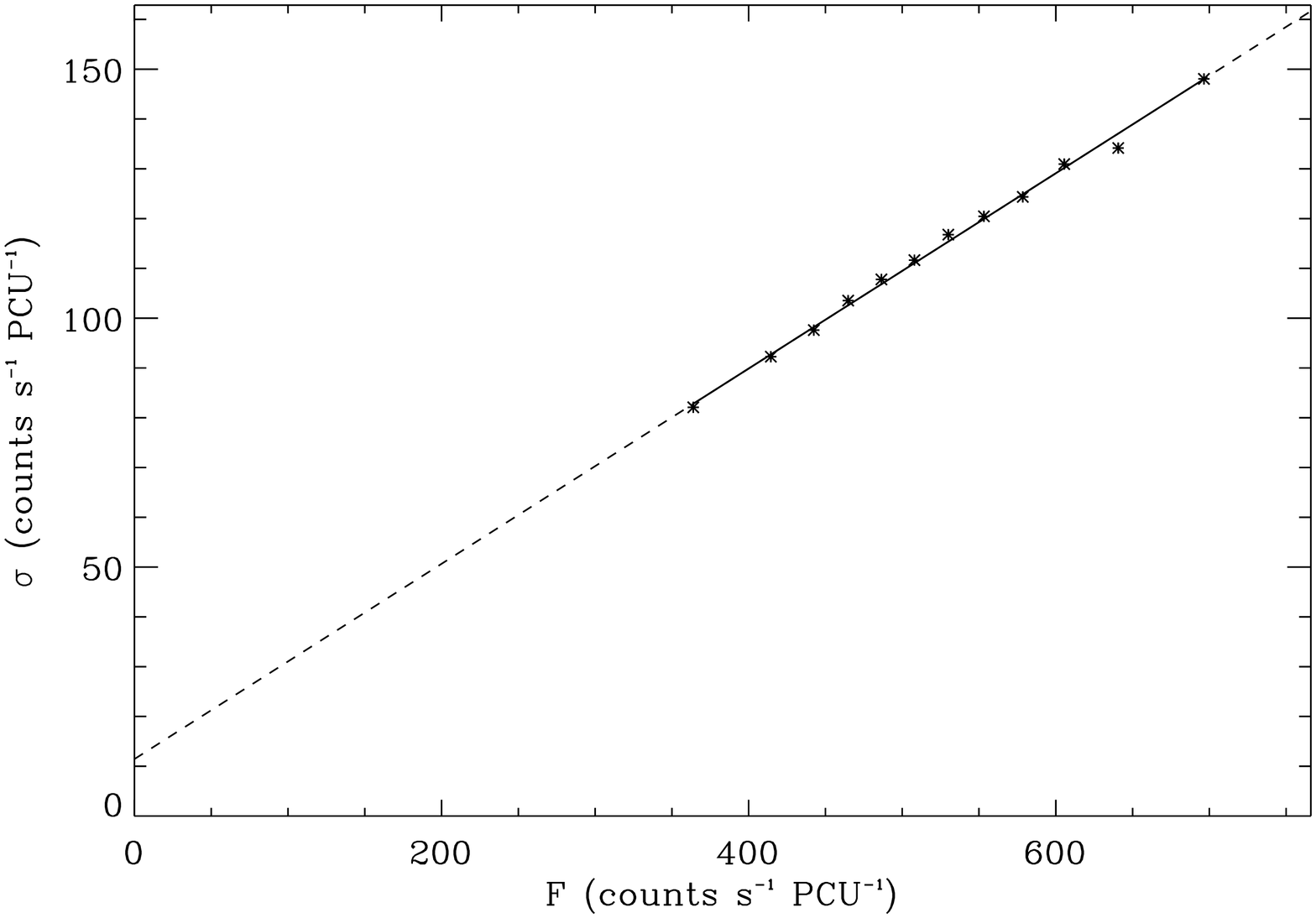}
\end{array}$
$
\begin{array}{lr}
	\includegraphics[width=5.5cm, angle=90]{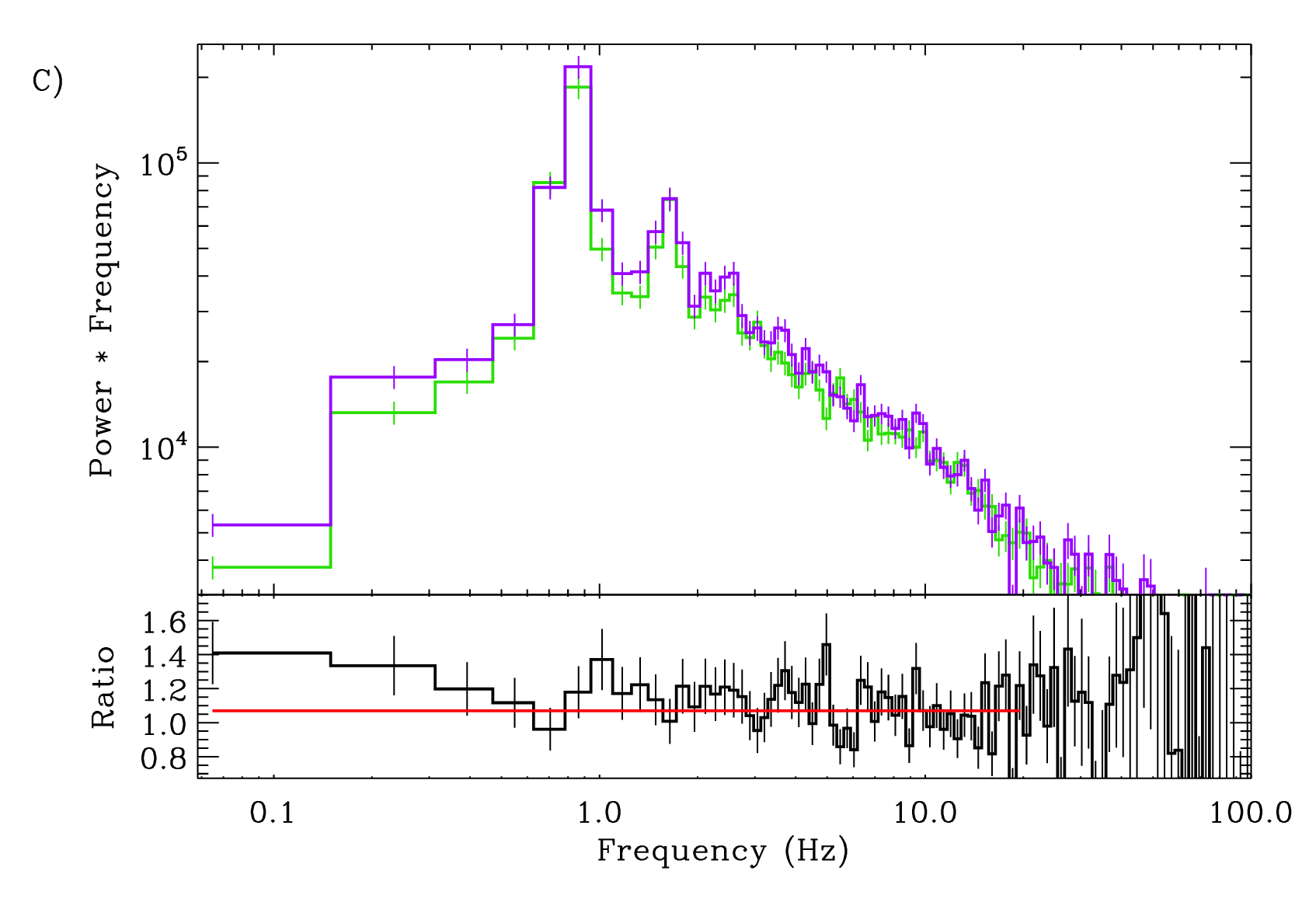} &
	\includegraphics[width=5.5cm, angle=90]{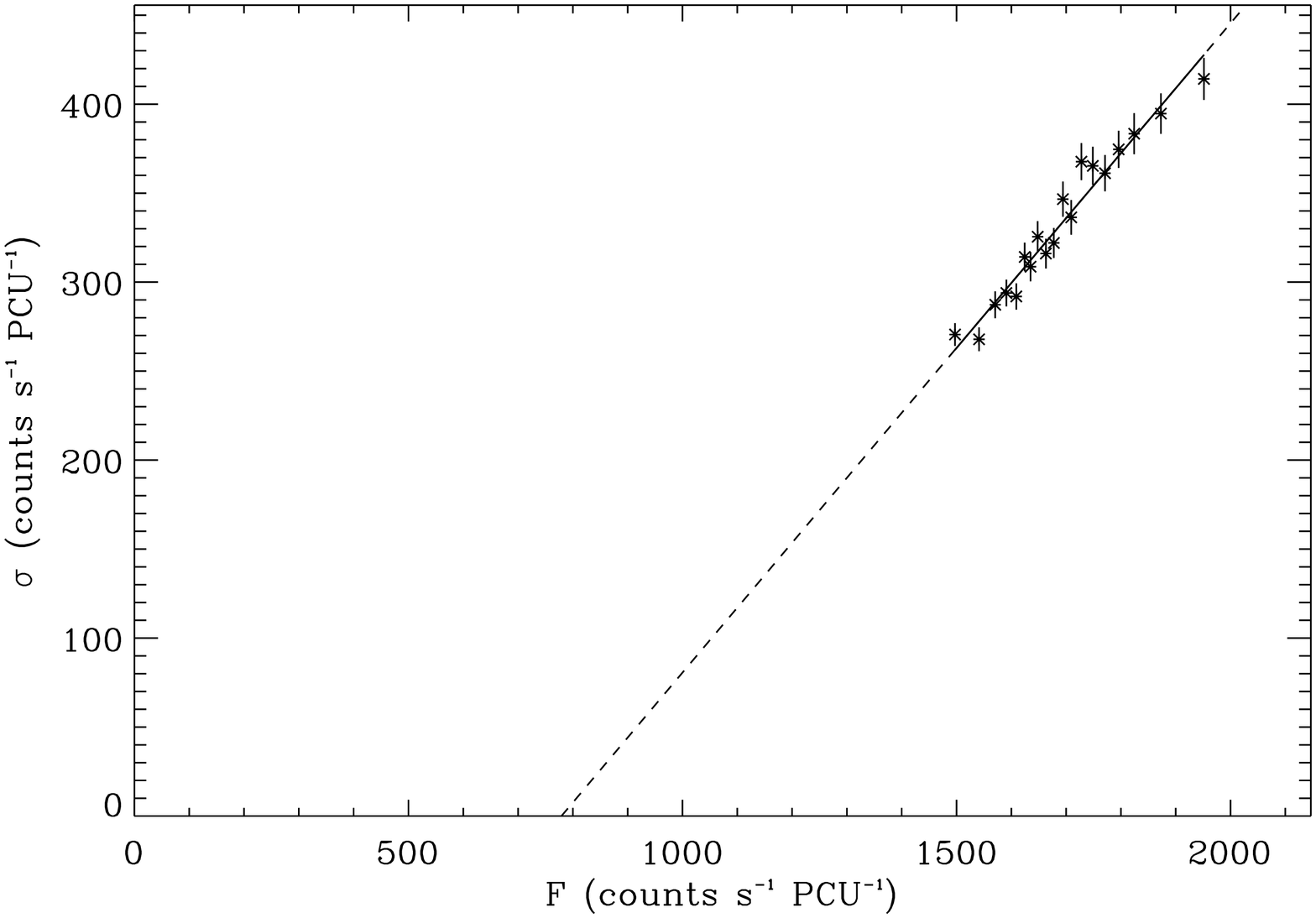}
\end{array}$
$
\begin{array}{lr}
	\includegraphics[width=5.5cm, angle=90]{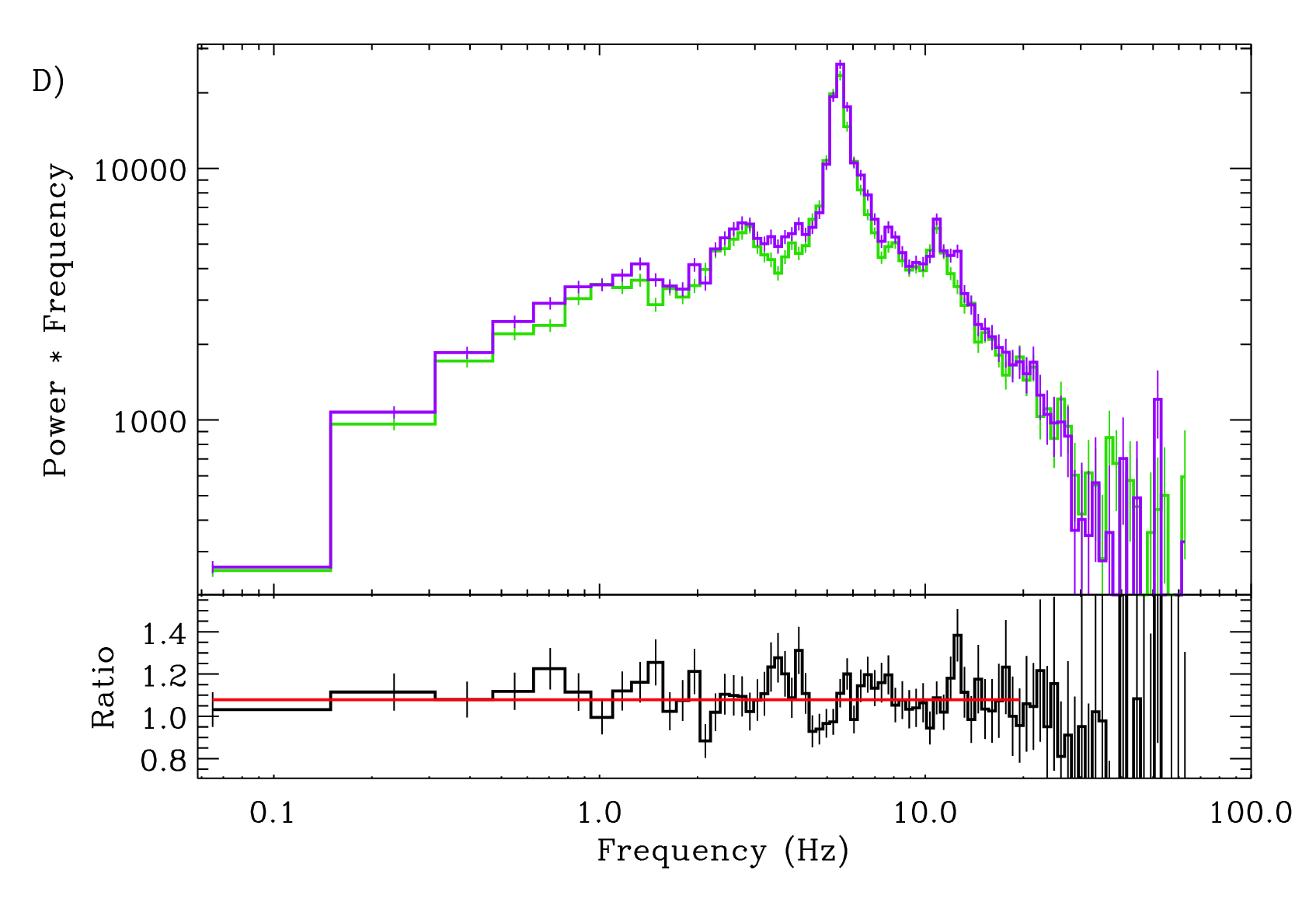} &
	\includegraphics[width=5.5cm, angle=90]{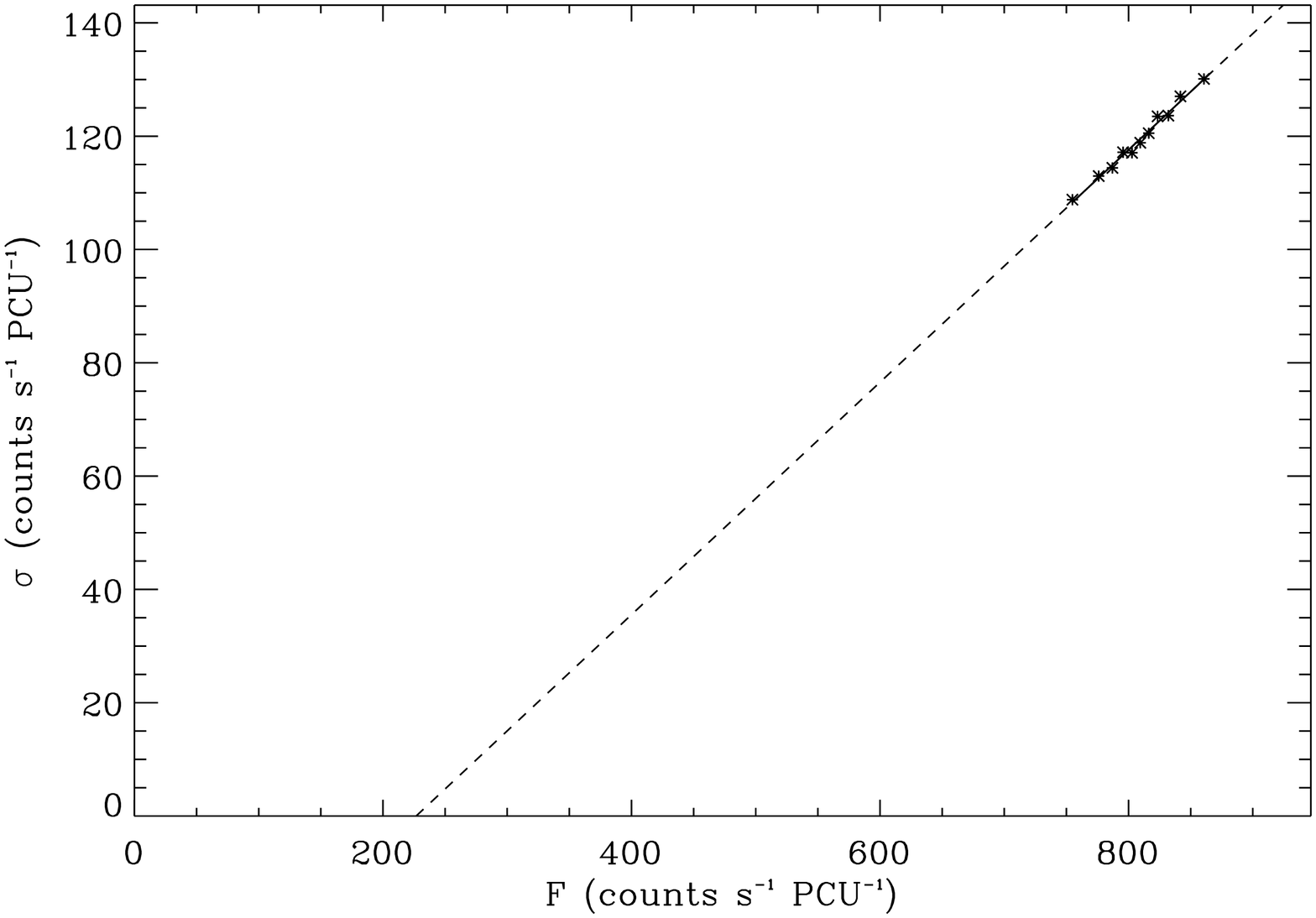}
\end{array}
$
 \end{center}
\end{figure*}

\begin{figure*}
\begin{center}$
\begin{array}{lr}
	\includegraphics[width=5.5cm, angle=90]{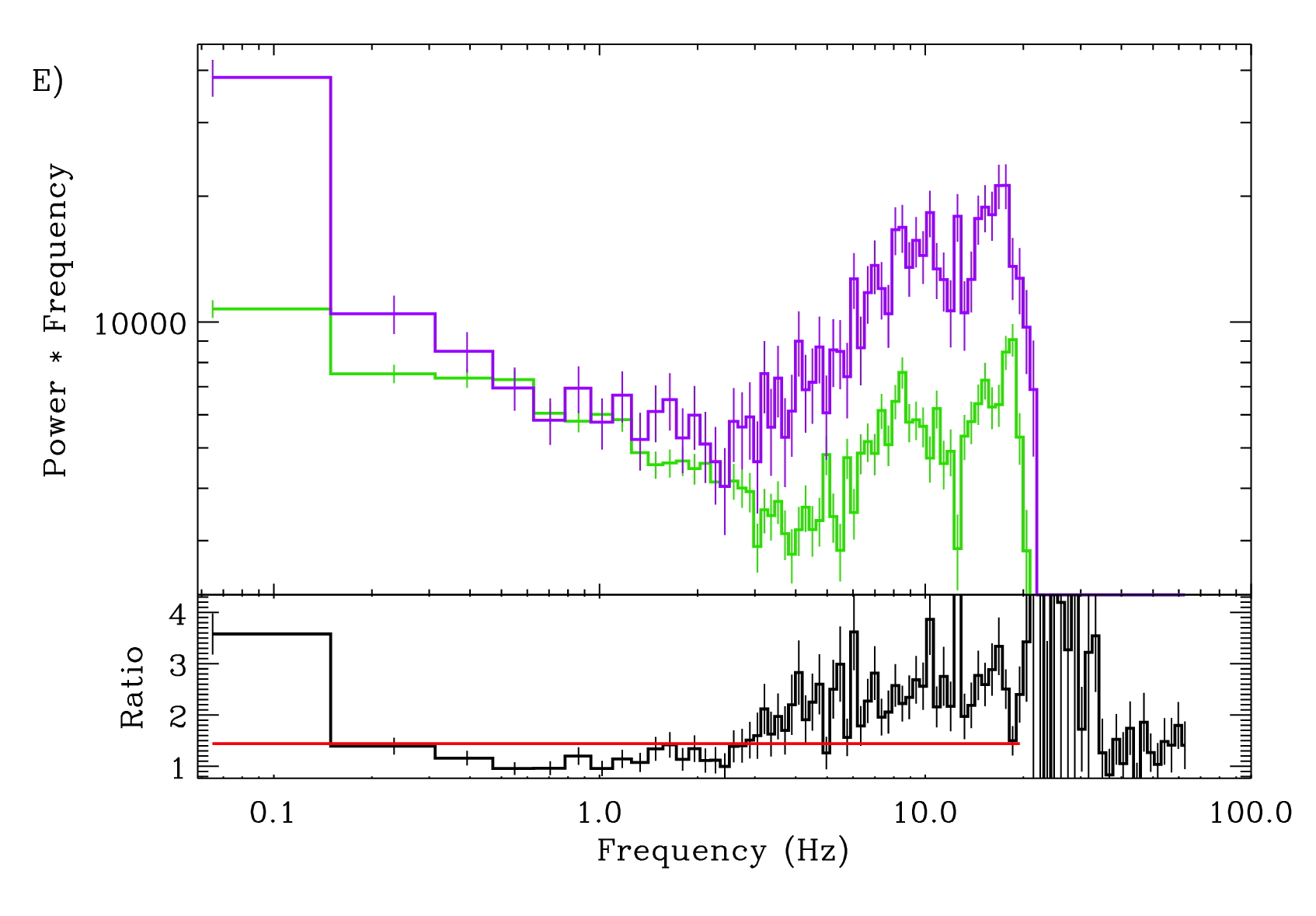} &
	\includegraphics[width=5.5cm, angle=90]{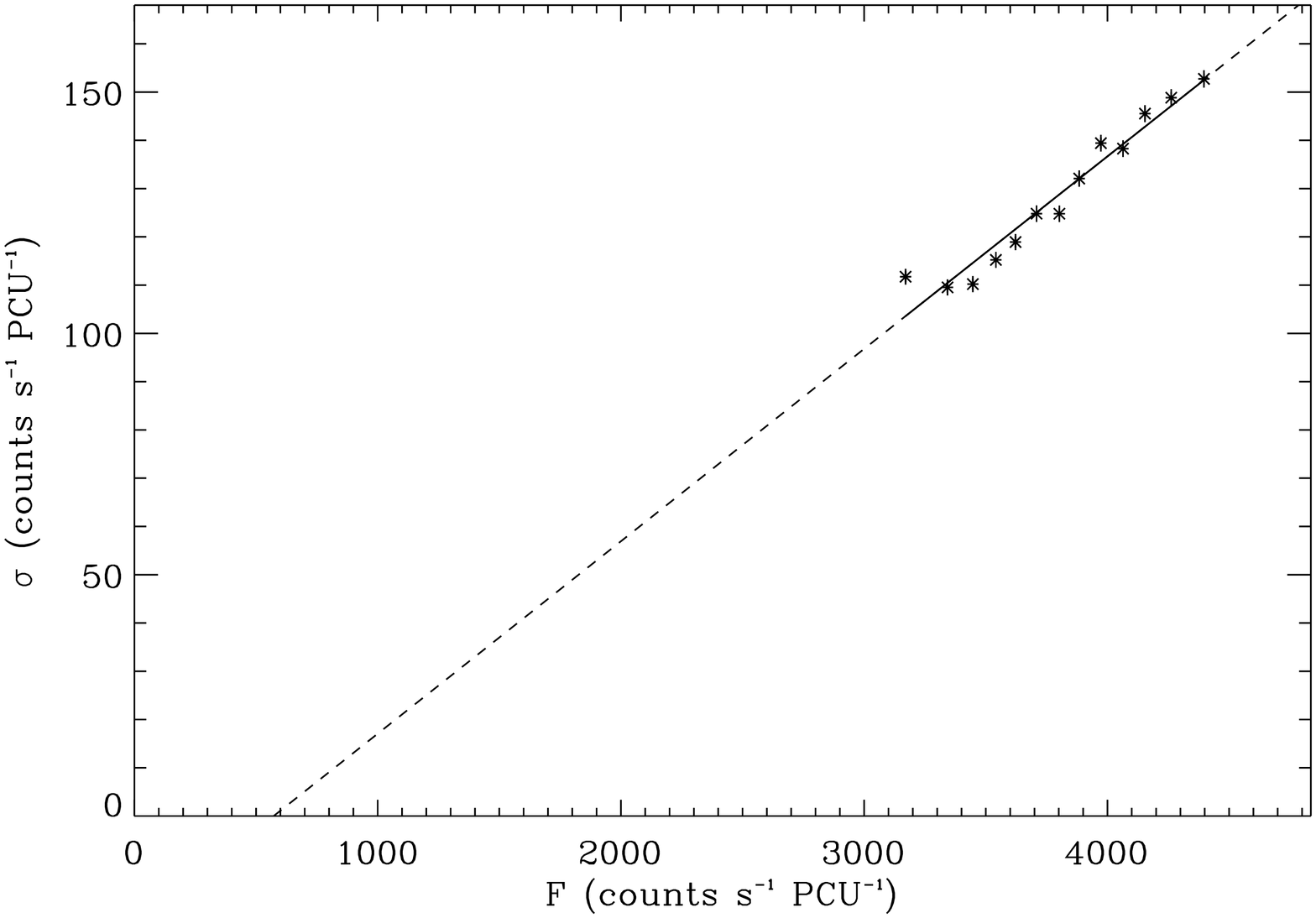}
\end{array}$
$
\begin{array}{lr}
	\includegraphics[width=5.5cm, angle=90]{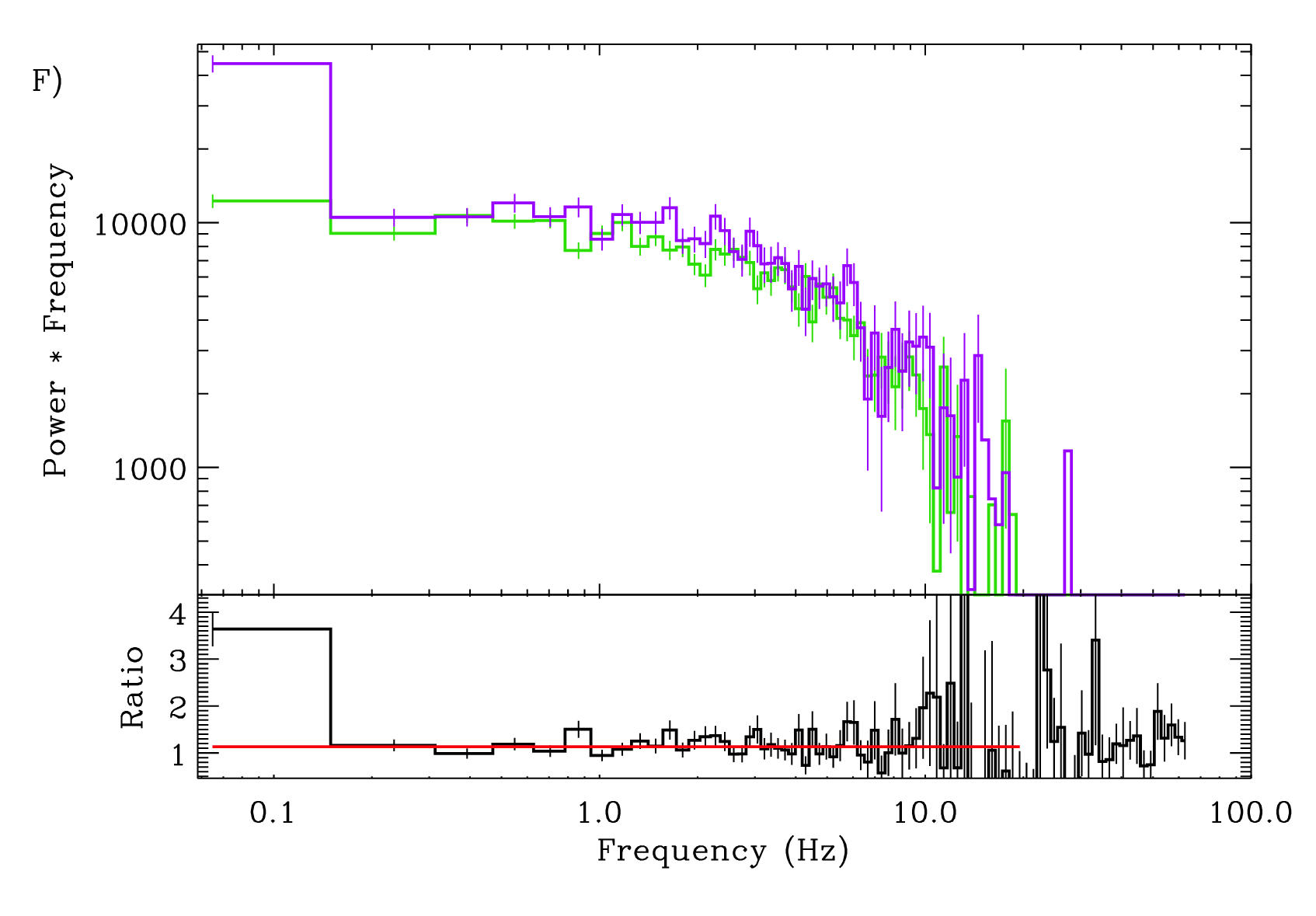} &
	\includegraphics[width=5.5cm, angle=90]{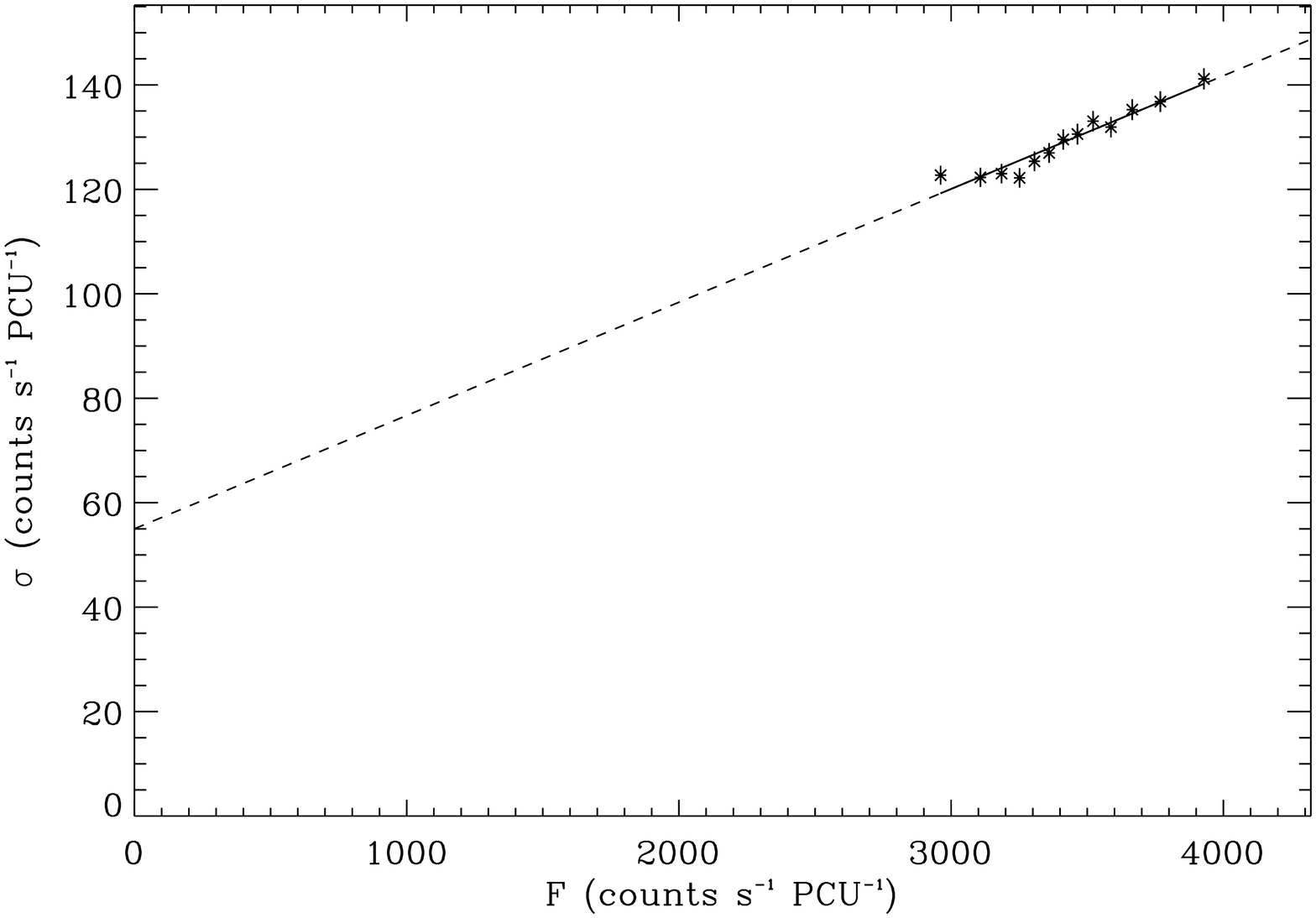}
\end{array}
$

\end{center}
\caption{Examples of noise-subtracted power spectra for the upper and lower flux quartiles and their corresponding rms-flux relations. These are normalised to the rms normalisation used to calculate the rms-flux relation, for this reason the flux dependence is still clearly visible in the power spectra. Low flux quartile power spectra are green, high flux are purple. \emph{Overleaf: A}: XTE J1550-564 (obs.id. 80135-01-04-00; Hard State) \emph{B}: GX339-4 (obs. id. 95409-01-12-01; Hard State), \emph{C}  XTE J1550-564 (obs. id. 30188-06-01-02; HIMS) \emph{D}: GX 339-4 (obs.id. 70109-04-01-01; HIMS) \emph{Above: E}: GRO J1655-40 (obs. id. 91702-01-63-00; Anomalous Soft State), \emph{F}  GRO J1655-40 (obs. id. 91702-01-62-00; Anomalous Soft State) }
\label{fig:goodrms}
\end{figure*}

\subsection{Calculation of the rms-flux relation}

The resultant background corrected lightcurves were then split into contiguous segments of 3 s in length and the mean count rate ($\langle F \rangle$) was found for each segment. Periodograms were then measured for the segments using absolute normalisation \citep[see e.g.][]{Uttley01}, in this normalisation the integrated area underneath the power spectrum is equal to the variance. The stochastic nature of the lightcurve results in scatter in the periodogram, the measured power values are scattered around the true value according to a $\chi^{2}$ distribution with two degrees of freedom \cite[see e.g.][]{vanderKlis89}. In order to reduce this effect and estimate the real value of the power in each frequency bin, the periodograms were averaged into flux bins. The number of periodograms per flux bin was dependent on the number of segments in the individual observations but ranged between 25 - 100, ensuring each observation contained a minimum of 10 flux bins. In order to find the rms, we first found the total variance in the required frequency range, given by summing the power over $\nu_1 - \nu_2$, and then multiplied by the frequency resolution ($\Delta f$) i.e. 

\begin{equation}
S = \sum_{j=1}^{W} \langle y_j \rangle  \Delta f 
\end{equation}

Where $S$ is variance, $\langle y_j \rangle$ is the power in a frequency bin with the bar indicating that this is averaged, and W is the number of periodogram points in the frequency range. In this paper the frequency range was chosen to focus on the region of the power spectra which shows power consistently in every state, whilst also displaying some of the most interesting features of the power spectra (spectral breaks and QPOs) -- 1-10 Hz. It is the convention with rms-flux relations \citep[see e.g.][]{Uttley01, Gleissner04, Uttley05, Gandhi09} to sample relatively restricted ($\sim$ 1 decade) frequency ranges for a number of reasons: Firstly, a lower limit of 1 Hz allows us to use shorter segments of lightcurve (in this case 3s), allowing us to measure the rms-flux relation on shorter timescales and to retrieve more flux bins from shorter observations. This means that a larger flux range is measured and the shape of the rms-flux relation is better defined. There is also a consistently high level of power at 1 Hz, particularly in the harder states \citep[see e.g.][]{Gierlinski08}, which again means we sample a wider range of fluxes. \cite{Gierlinski08} show that the broad band noise displays little evolution within this frequency band, meaning that the stationarity in PSD shape required to measure the short-term rms-flux relation is more likely to be observed -- this excludes certain observations with QPOs within this frequency range, which may display different behaviour to the underlying noise \citep[see e.g.][]{Heil11, Rodriguez11}. Figure \ref{fig:goodrms} shows that although there is often some power above $\sim$10 Hz, in general this is insignificant when compared to that observed within the chosen 1-10 Hz range.

As the power in each frequency bin contains a contribution from the noise this needs to be estimated and removed so that the rms can be found $\hat{\sigma} = \sqrt{S - P_N W \Delta f}$, where $P_N$ is the power density for the Poisson noise component of the variance and is given by $P_N = 2 (x + B)$, $x$ is the source count rate and $B$ is the background. The flux binning allowed for an accurate calculation of the errors on the binned $\hat{\sigma}$ values, these were found in a more general manner to that applied in \cite{Gleissner04}. See appendix A for further discussion of this error.

\subsection{Fitting the linear relation}

Following \cite{Uttley01} we have assumed that the rms-flux relation is linear and described by 
\begin{equation}
	\hat{\sigma} = k (F-C_{x})
\end{equation}
Where the gradient of the relation is given by $k$, and $C_{x}$ is the intercept on the x-axis. Non-zero intercepts have been widely observed previously. \cite{Gleissner04} found both positive and negative x-axis intercepts in the rms-flux observations from Cygnus X-1, possibly indicating the presence of two variable components within the light-curve. 


\subsection{Testing for ``good'' rms-flux relations}
\label{subsec:tau}

In order to evaluate the rms-flux relation in each individual observation we employed a method similar to that used in \cite{Gleissner04}. Testing for correlation using Kendall's rank correlation coefficient, $\tau$ \citep{Press92}, in addition to the $\chi^{2}$ test to evaluate the fit. Kendall's tau tests for correlation which the bulk of points follow without being heavily affected by outliers. Due to the lower levels of variability and count rates in data from the BHBs compared to observations of Cygnus X-1, less stringent limits were used to define ``good'' data than those used in \cite{Gleissner04}. Whilst their minimum threshold was set at $\tau_{min}$ = 0.9 we have used $\tau_{min} = 0.5$. Observations with this value of $\tau$ still display a positive linear correlation but greater scatter is allowed around the trend. The assumption was made that if $\tau$ was significant at a level of 2$\sigma_{\tau}$ the data were sufficiently strongly correlated. In addition to this we require the fit to the linear model to be accepted using $\alpha~<$ 0.003 (i.e. not rejected at 3$\sigma_{\chi^2}$).


In general the data were found to agree with the assumed linear model, with the majority of discrepant observations due to data at low count rates and/or with low fractional rms. Assuming a background rate of approximately 7 ct/s/PCU and a count rate of 100 ct/s/PCU the fractional rms of the Poisson noise level is 2.14$\%$. Although this level drops rapidly as the count rate increases, a significant amount of power above the noise is required to satisfactorily measure the rms. Some observations with count rates below $\sim$ 500 ct/s/PCU and a fractional rms of $\sim$3$\%$ (where the noise level drops to 0.4$\%$) may only just have enough power for the source variance to be measured accurately in the observed frequency band. For this reason a limit of 3$\%$ fractional rms was imposed, observations where the fractional rms is less than this within the frequency band were discarded. Most observations with count rates lower than 100 ct/s/PCU are typically observed in a hard state where the fractional rms is higher than 10$\%$ \citep[see e.g.][]{Remillard06} and so are less strongly affected by this limit. However, in a standard soft state the variability often drops drop below the required 3$\%$ fractional rms level, so even though the count rate is high few observations of sources in this state are included in the following analysis.

Many of the other rejected observations had low count rates, resulting in flux binned PSDs with low signal to noise and negative Poisson noise-subtracted variances. Observations containing negative variance bins were therefore excluded from the sample. Short observations also limit the signal to noise for the flux binned PSDs as the number of rms points per flux bin are low. The length of good time required for a detected rms-flux relation varied according to the count rate and the level of intrinsic variability within the observation, but more than 1 ks was generally required. 

\section{Results}

\subsection{Ubiquity of the rms-flux relation}

1,961 observations of the sources listed in table \ref{tab:sources} have been tested for a simple linear rms-flux relation. Of these observations 1105 displayed the 3$\%$ fractional rms required for the rms-flux relation to be measured. A "good'' rms-flux relation as defined by the criteria set out in section \ref{subsec:tau} was observed in 634 of these observations. 

Observations where a positive rms-flux correlation was not found were inspected individually. Non-stationarity of the power spectrum within an observation can cause a `wavy' rather than a linear rms-flux relation to be observed, such as those shown in figure 8 of \cite{Gleissner04}. To some extent a further example is shown in Figure \ref{fig:goodrms} panel E, there is a lower flux segment of time series within this observation where the rms-flux relation is slightly flatter, for this reason the lowest flux bin has a slightly higher rms when compared to the other bins. The flux-binned PSDs of long, bright observations which deviated significantly from a positive linear rms-flux correlation were inspected for any shape change over the course of the observation. In some observations plotting the individual flux binned power spectra for each point in the rms-flux relation, revealed QPOs shifting with flux in either frequency or amplitude over the course of the observation, similar to those discussed in \cite{Heil11} and \cite{Rodriguez11}. In other cases it was apparent that the shape of the whole power spectrum was evolving with flux. No examples of non-linear or negatively correlated rms-flux relations were found within the sample that could not be explained by one of the above reasons. Where sufficient data are available and the variability is stationary, the rms-flux relation appears to be ubiquitous in the broad-band noise from BHBs.  


Figure \ref{fig:goodrms} illustrates observations of the rms-flux relation across a range of different states and sources, from the low-hard to an anomalous soft state.


\begin{figure}
\begin{center}
	\includegraphics[width=6.2cm, angle=90]{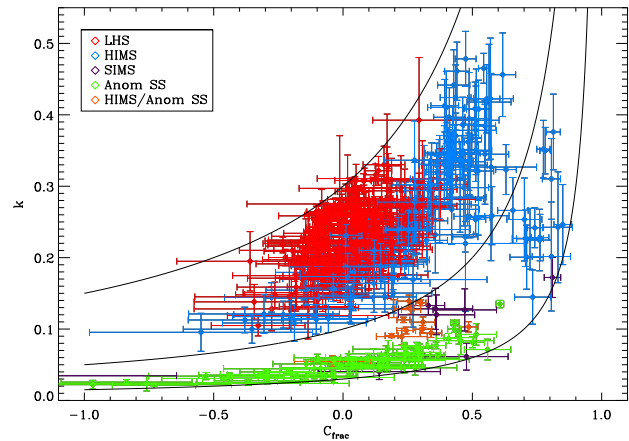}
\end{center}
\caption{Results for all good rms-flux relations. The colours correspond to the state of the source and states are defined as follows: \emph{LHS} - Low hard state; \emph{HIMS} - Hard Intermediate State; \emph{SIMS} - Soft Intermediate State; \emph{Anom SS} - Anomalous Soft State; \emph{HIMS/Anom SS} - observations with characteristics of both the HIMS and the Anomalous Soft State which are uniquely observed in H 1743-322. The contours shown are lines of fractional rms at 30$\%$, 10$\%$ and 3$\%$.}
\label{fig:allkC}
\end{figure}

\begin{figure}
\begin{center}
	\includegraphics[width=6.2cm, angle=90]{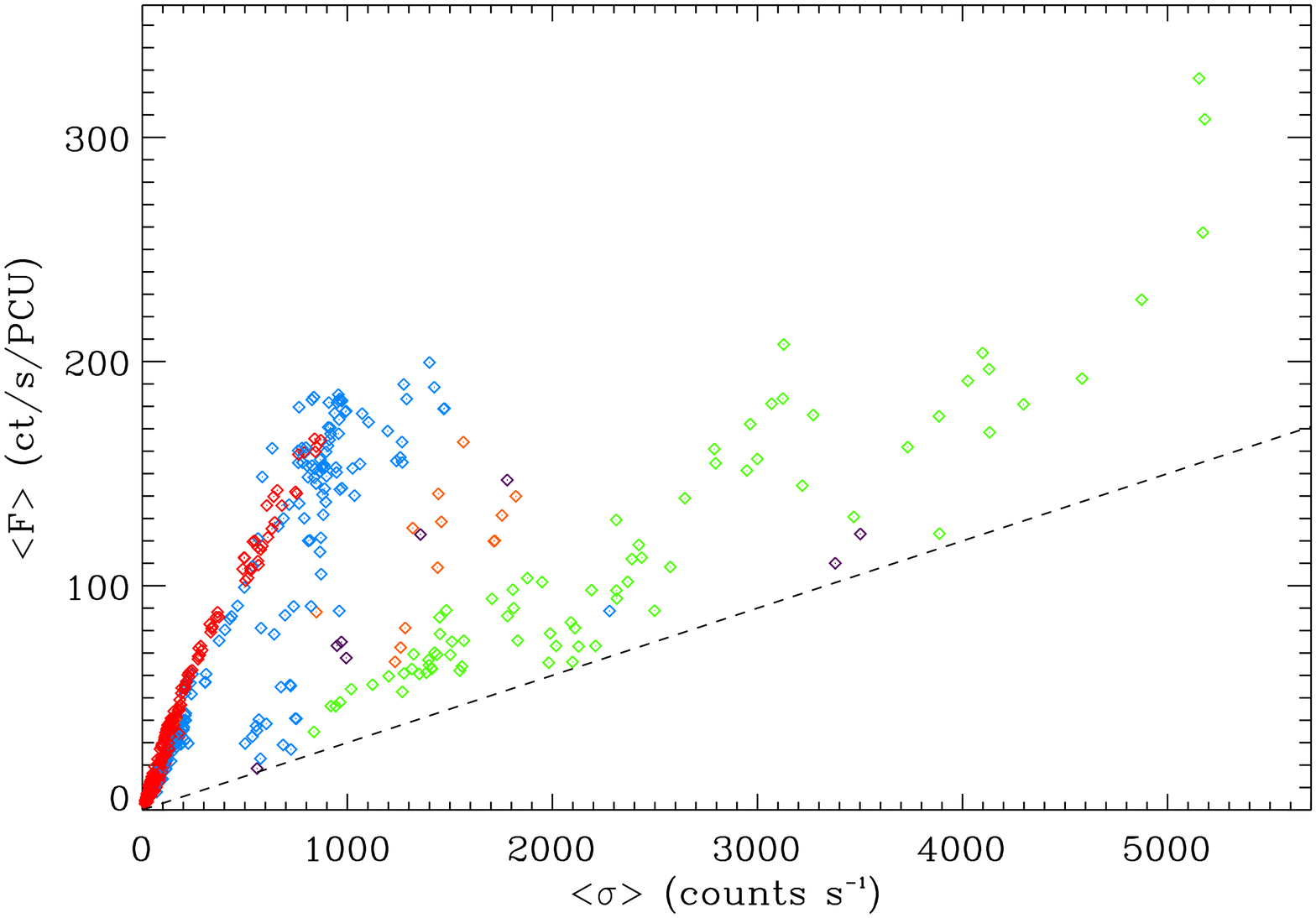} 
	\includegraphics[width=6.2cm, angle=90]{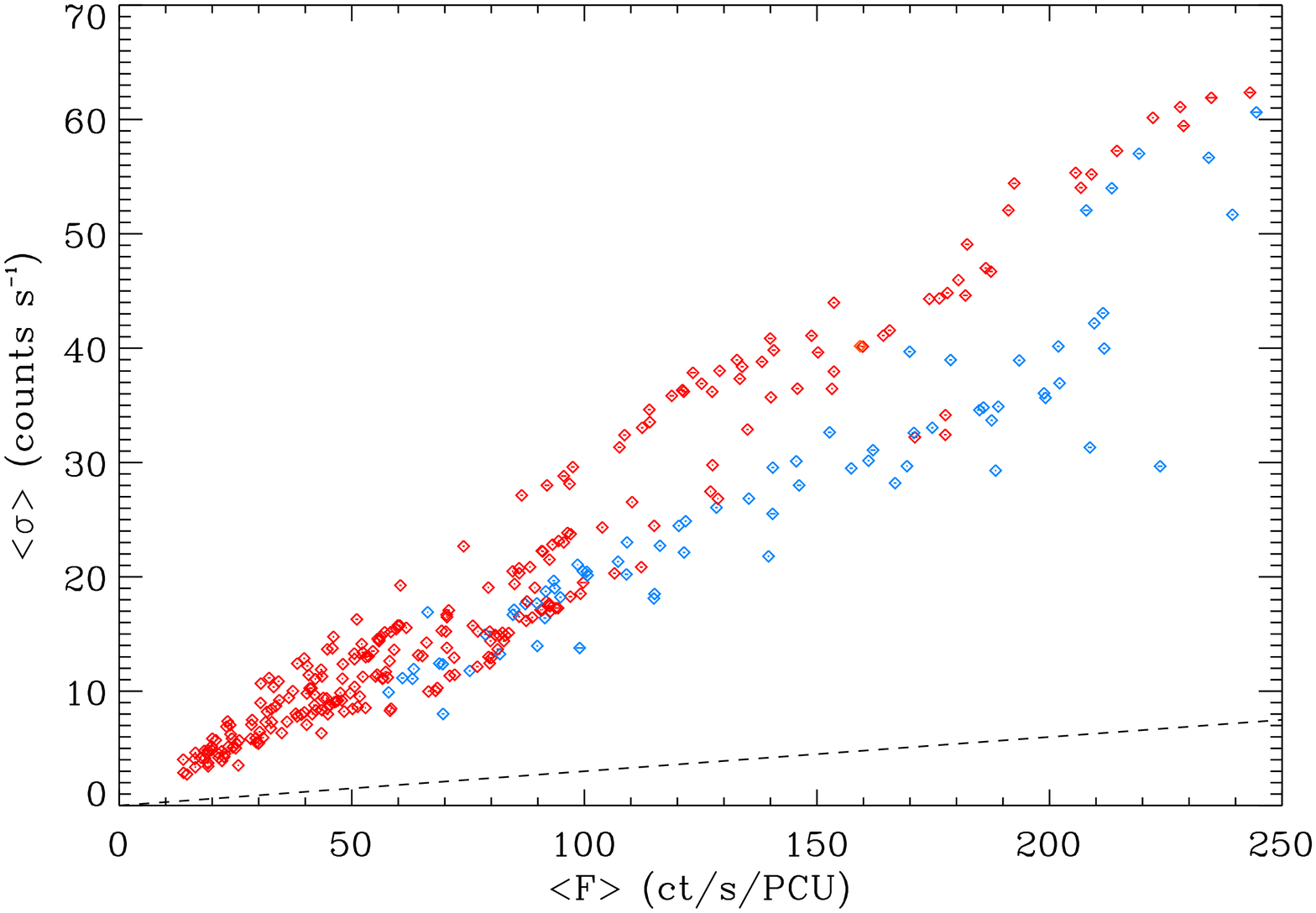}
\end{center}
\caption{ \emph{Top.} Long term rms-flux results for all good rms-flux relations in the 1-10 Hz range. \emph{Bottom} As \emph{Top} but concentrating on lower values. The dotted line plots the lower limit on detection for the rms-flux relation -- 3$\%$ fractional rms. The colours correspond to those used for Figure \ref{fig:allkC}.}
\label{fig:alllongrms}
\end{figure}

Figure \ref{fig:allkC} shows the gradients against the intercepts for all sources within the sample. As the flux-intercept ($C_{x}$) is dependent upon source count rate the values plotted are shown as a fraction of the mean count rate over the observation (the fractional intercept: $C_{x}/ \langle F \rangle = C_{frac}$), this allows the parameters from all sources in all states to be compared directly. The ``banana'' shaped paths within this plot are explained as lines of fractional rms if we re-write the gradient as 
	\begin{equation}
	k = \frac{\langle \hat{\sigma} \rangle}{\langle F \rangle - C_{x}} 
	\end{equation}
In this case the value of the gradient may be interpreted as the fractional rms of the observation after removal of the intercept on the x-axis, if $C_{x} = 0$ then $k = \langle \hat{\sigma}_{frac} \rangle$ (where $\langle \hat{\sigma}_{frac} \rangle  = \langle \hat{\sigma} \rangle /\langle F \rangle$ and is fractional rms). This illustrates the importance of the intercept in interpreting these results. Observations with the same overall fractional rms and count rate but different intercepts ($C_x$) will show rms-flux relations that pivot around a common point, meaning either steep or shallow gradients may arise from observations with the same mean rms and flux. When $k$ is plotted against $C_{frac}$ then the changes in $C_{x}$ and $\langle F \rangle$ work to produce the observed curves: $k \rightarrow \infty$ as $C_{frac} \rightarrow 1$ and $k \rightarrow 0$ as $C_{frac} \rightarrow -\infty$. The lower and further to the right the curve on Figure \ref{fig:allkC}, the lower the fractional rms of the observations. As fractional rms is well known to be related to source state \citep[see e.g.][]{McClintock06} then the position of points on the $k$ vs. $C_{frac}$ diagram provides a link to the energy spectral properties of the observation.

Whilst the paths on this plot are therefore just contours of different fractional rms values, it clearly displays the full range of intercepts and gradients seen within the sample. It is also apparent that $C_{x}$ is less well constrained at negative values than positive ones, this is partly an artefact of the fitting, by nature negative intercepts are further from the actual observed rms-flux values than positive ones and therefore the point where the fitted line intercepts the x-axis is less certain. The range of intercepts and gradients observed within the sample is striking, particularly when compared to those observed by \cite{Gleissner04} from Cygnus X-1. Whereas their figure 5 indicated only a few observations with negative flux-intercepts in the softer states and none in hard states, we observe a wide distribution of positive and negative $C_{x}$ values over a range of fractional rms values. Section \ref{sec:statedep} explores the positions of points in Figure \ref{fig:allkC} in relation to source state.

\subsection{Long term rms-flux relation}

In a similar manner to \cite{Gleissner04} the long-term rms-flux relation is plotted for all of the sources in the sample by taking the mean rms and flux values for each observation in the 1-10 Hz band. The rms-flux relation measured within each observation relies on the PSD remaining the same shape in the 1-10 Hz frequency band. The power spectra in intermediate states can show dramatic changes in between observations, with QPOs and broad band components changing in both frequency and amplitude. On longer timescales there is not the equivalent stationarity in power spectral shapes observed within a single observation, therefore the long term relationship in these states is not representative of that observed on short timescales. In contrast evolution of the power spectra within the hard state occurs mainly below the 1-10 Hz range and although the shape does not remain completely stationary it is more suitable for comparison with the short term rms-flux relation. 


 In Cygnus X-1 separate linear rms-flux relations were observed in the long-term relations in both the hard and softer states. Figure \ref{fig:alllongrms} shows the mean rms and flux values for all good rms-flux observations within the sample. There are two clear paths, the upper one is that followed by sources in the hard state, the lower shows those in anomalous soft states. Intermediate states fall between these two lines. The lower edge of this plot is artificially introduced by the requirement that there be at least 3$\%$ fractional rms within the 1-10 Hz frequency band (marked by the dotted line), so there is a wider range in the lower fractional rms soft state observations than appears here. However, the relations for individual outbursts appear to be remarkably linear over a wide range of count rates, and there is distinct similarity in those observed between different sources. The agreement in these paths over repeated outbursts of GX 339-4 has been remarked upon in \cite{Munoz10} and we return to this point in section \ref{sec:statedep}. This plot also clearly shows the sharp drop in fractional rms over a few observations within the transition from the HIMS through the SIMS to soft anomalous states, seen in the points which lie in between the hard (upper) and soft (lower) state rms-flux relations. The lack of intermediate points at high count rates occurs because these observations are taken during bright flares where no state changes were observed. Further differences between the relations over the course of the outbursts are discussed in the following section.  





\subsection{Evolution of rms-flux relationship with BHB state}
\label{sec:statedep}

\begin{figure*}
\begin{center}

	\includegraphics[width=5.5cm, angle=90]{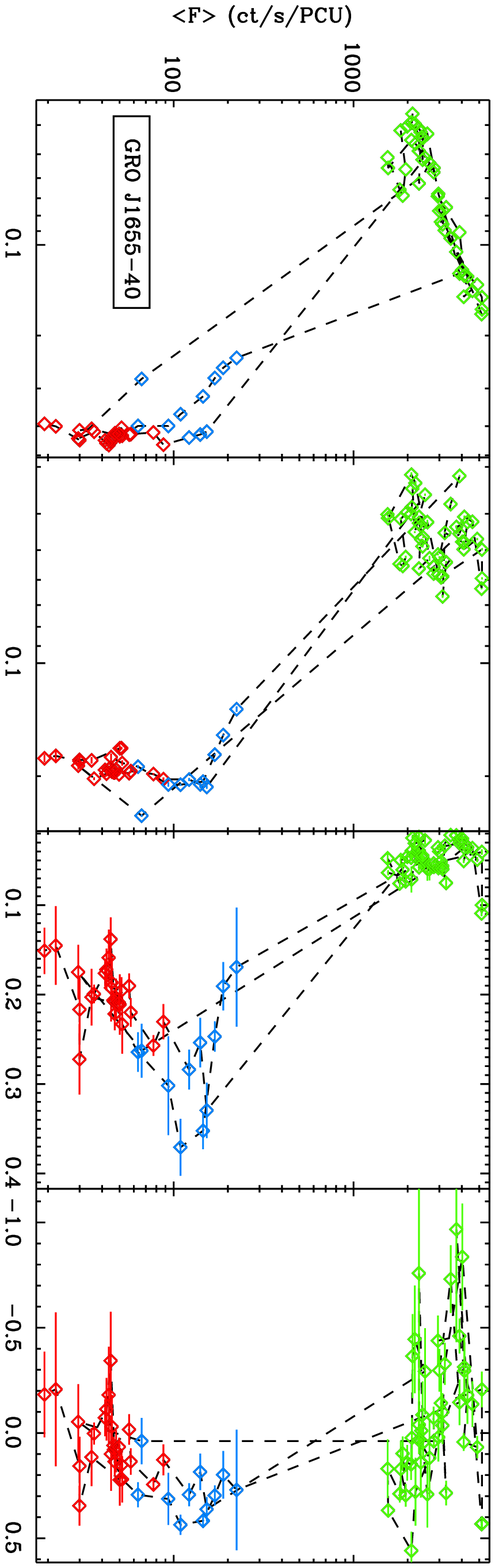}

	\includegraphics[width=5.5cm, angle=90]{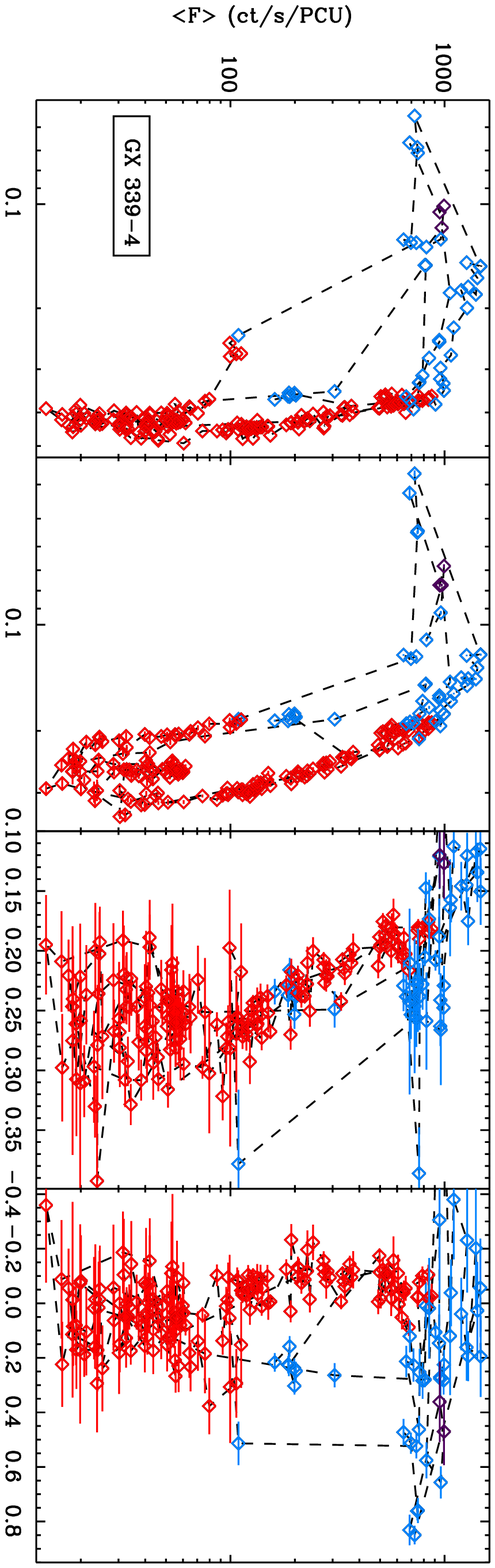}

	\includegraphics[width=5.5cm, angle=90]{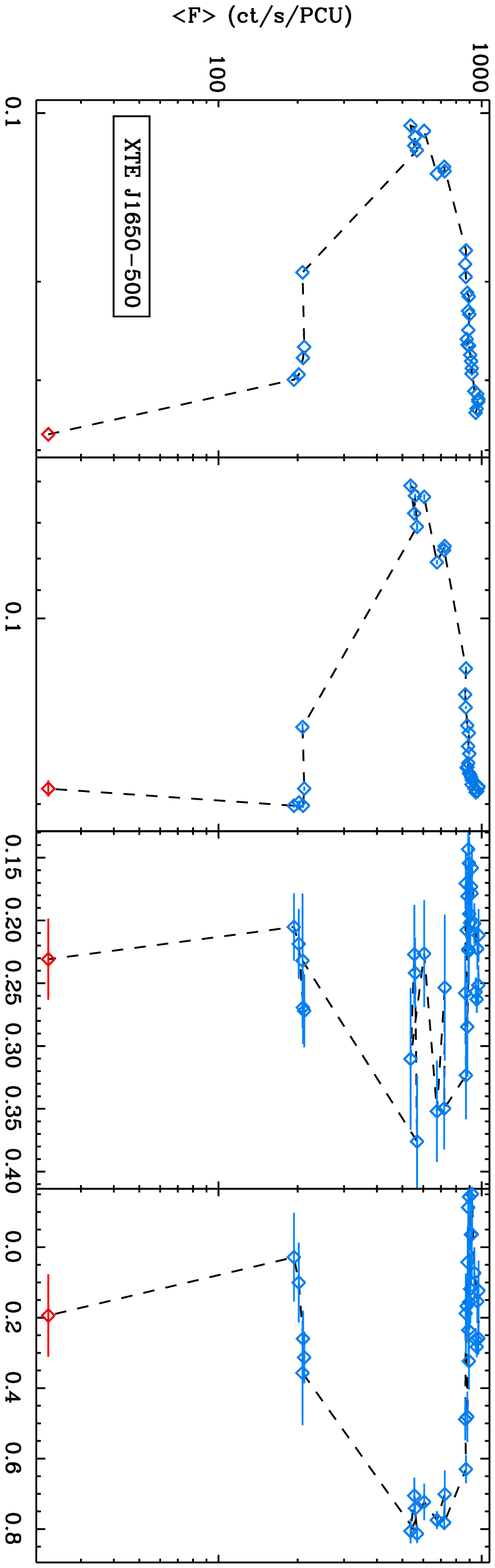}

	\includegraphics[width=5.9cm, angle=90]{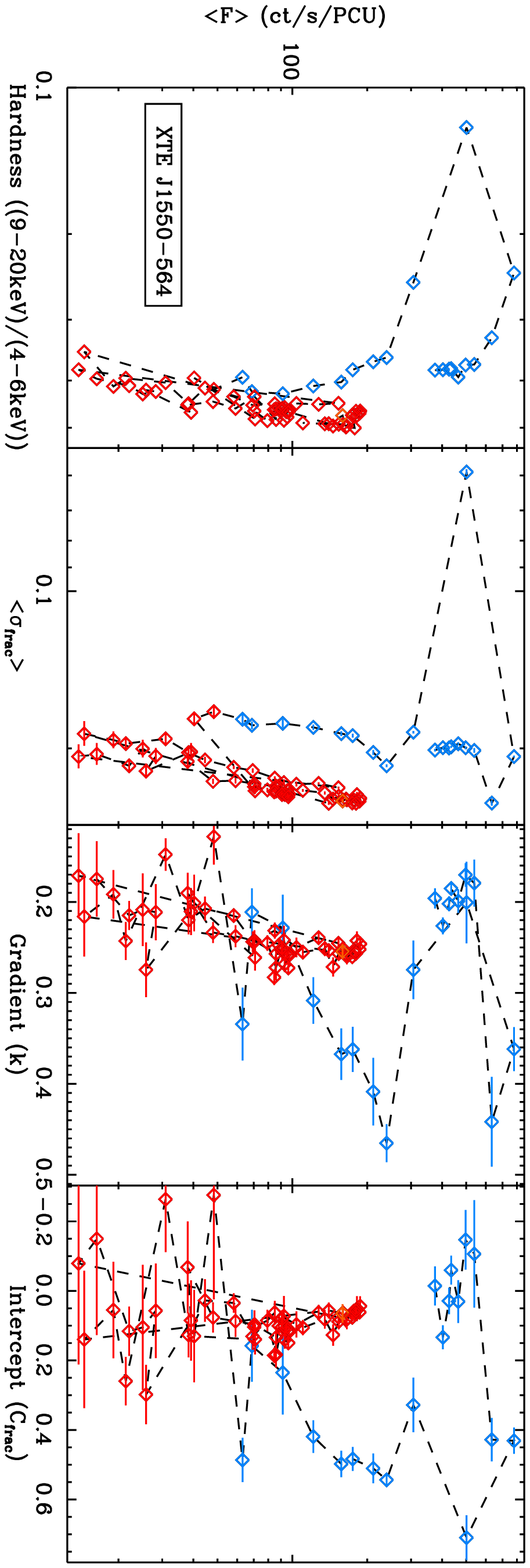} 





\end{center}
\caption[]{Hardness ratio, fractional rms, Gradient and Intercept vs. Count rate for four sources. Points below 3$\%$ fractional rms are excluded from the analysis. Colours correspond to those used in Figure \ref{fig:allkC}.}
\label{fig:Intplts}
\end{figure*}

\cite{Gleissner04} noted a distinct difference between the behaviour of $k$ and $C_{x}$ in different states of Cygnus X-1. In the softer states $k$ was generally found to be lower and $C_{x}$ had a much wider range of values than those observed in the hard state. 

Figure \ref{fig:Intplts} shows the evolution of the hardness ratio, fractional rms, gradient ($k$) and intercept ($C_{frac}$) for sources with the largest number of good observations. The flux-intercept is again shown as a fraction of the mean count rate, allowing for comparison between different sources and different states of the same source. All four parameters plotted are dimensionless.

The state of a source is usually identified based on a combination of
its luminosity, timing properties (the power spectrum) and
energy spectrum. The most important single diagram
currently in use for diagnosing and tracking the evolution
of states is the hardness intensity diagram (HID).
The outbursts of transient black holes, and some neutron star and cataclysmic variable systems tend to follow a similar ``q" shape path around this diagram \citep{Homan01, Belloni05, Belloni10, Kording07} beginning in a low/hard state (HS), evolving through hard then
soft intermediate states (HIMS/SIMS) to the soft (thermal
dominated) state, before changing back to the hard state
via lower luminosity intermediate states.
The HID itself is a powerful tool based
on only the source intensity (in a particular X-ray band)
and a very crude indicator of the energy spectral shape,
with no explicit dependence on timing properties.
But as discussed in \cite{McClintock06} and  \cite{Belloni10} the timing properties of these objects evolve as clearly, if not more clearly than the energy spectral properties do. For more details see \cite{Belloni10, Dunn10} and references therein.

It is already well-known that the fractional rms (in the
usual $2$--$20$ keV X-ray band) is lower in the soft
than hard state. \cite{Munoz10} presented a new plot which demonstrated how the pattern described by the HID is repeated by the rms. In the second panels of Figure \ref{fig:Intplts} we show
an altered version of this new diagram -- the rms intensity diagram (RID) -- the
natural variability
counterpart to the HID, which shows the intensity
against fractional rms amplitude as discussed in \cite{Munoz10}. There is a remarkable similarity between the HID
and RID. The RID is of course very similar to the long-term
rms-flux plot
in Figure \ref{fig:alllongrms}, indeed it is simply a
geometric transformation of that plot and contains the
same information.
\cite{Munoz10} clearly demonstrate that the position
on the RID is related directly to the source state in GX 339-4.
This demonstrates that variability is every bit as
strong an indicator (or diagnostic) of outburst evolution
and states as is the energy spectrum. 

 For GX 339-4, XTE J1550-564 and XTE J1650-500 the classic `q' shape of the hysteresis is replicated by the fractional intercept over the course of the outburst. The intercept appears close to zero in the hard state, but becomes strongly positive as the source transitions into the HIMS before moving back again. This can be seen in Figure \ref{fig:allkC}, the red and blue points (HS and HIMS respectively) occupy a similar area within the plot (which is expected as one evolves from the other) but the highest values of C$_{frac}$ occur when the source is in the HIMS. This behaviour is seen in at least 3 repeated outbursts of GX 339-4 which all follow remarkably similar paths in both the hardness and rms - intensity diagrams. In contrast the high flux observations of GRO J1655-40 show a wide range of intercepts including ones which are strongly negative. In this case the power spectra indicate that the source is in an anomalous soft state \citep{Belloni10}, suggesting that the behaviour of both the intercept and gradient is strongly state related (these observations correspond to some of the green points in Figure \ref{fig:allkC}). The observations of GRO J1655-40 in the hard state and HIMS still display similar strong positive intercepts observed in other sources. This can be observed in the lower flux ($<$ 500 ct/s/PCU) observations in Figure \ref{fig:Intplts}. The appearance of intercepts close to zero in the hard state rising to strong positive values in the HIMS and then displaying a wide range of values, including strong negatives, in the soft states is replicated for other sources in the sample (i.e. H1743-322).

The state dependence of the intercepts, explains the wide range of values in Figure \ref{fig:allkC} at low fractional rms, and almost bimodal spread between those with high and low k values. Observations with C$_{frac}~>~0.5$ and k $>$ 0.1 are from a hard-intermediate state (blue), whereas when C$_{frac}~<~0.5$ and k $<$ 0.1 the source is in an anomalous soft state (green). Observations in the SIMS have a range of values but this is likely to be because they are strongly influenced by the behaviour of the QPOs \citep[see][]{Rodriguez11}. Figure \ref{fig:allkC} demonstrates that both k and C$_{frac}$ are not only dependent on the fractional rms within the 1-10 Hz range, but also upon the state of the source.

\subsection{Flux dependence of the PSD shape}
\label{sec:freqdep}




In order to investigate any possible flux dependence of the PSD, power spectra were measured for 100 s long segments of lightcurve and grouped into four flux bins. The averaged PSDs from the highest flux bin were then compared with that from the lowest. Figure \ref{fig:goodrms} shows the results for the six example observations. We tested the constancy of the PSD shape by fitting the ratio of high and low flux PSDs with a constant, using the $\chi^{2}$ statistic to evaluate the quality of the fit. The ratio was fitted up to 20.0 Hz, above this point some observations are dominated by Poisson noise (see e.g Figure \ref{fig:goodrms}f).  If the fit was judged to be poor ($>$ 3$\sigma_{\chi^{2}}$ ) then the observations were subject to further investigation. 

Little evidence for flux dependence was found over the observed frequency range in most of the observations: in total only 20 of the good observations were identified as having obvious frequency-dependent differences between the PSDs from the upper and lower flux quartiles. Different binning was tested in order to eliminate possible spurious results  caused by sharp features (e.g. QPOs) appearing shifted in frequency between the two power spectra. Fifteen of these observations were from the 2005 outburst of GRO J1655-40 whilst it was in the anomalous soft state. The power spectra all show strong low frequency noise with a high frequency broad band component, an example of which is shown in Figure \ref{fig:goodrms}e. The two flux binned power spectra indicate that the peaked band-limited component shows a greater change in rms with flux than the power law. The observations with clear flux dependence are not consecutive and the appearance of the power spectrum changes between this and the typical soft state shape. The power spectra in Figures \ref{fig:goodrms}e and f were taken almost exactly a day apart and show distinctly different behaviour, although the fractional rms in the 1-10 Hz band remains similar (3.7 $\%$ vs. 3.5 $\%$). Figure \ref{fig:goodrms}f is typical of that observed in the soft state, only with a higher fractional rms, whereas Figure \ref{fig:goodrms}e is more ``anomalous''.

The upper and lower flux binned power spectra in Figure \ref{fig:goodrms}e have been separately fitted using {\sc xspec} to test if the change in power spectral shape can be mainly explained by the strength of the peaked noise components. The model used a bending power, plus two Lorentzians for the high frequency broad band component and a separate narrow Lorentzian to model the very low frequency component, which can be seen in the lowest frequency bin of Figure \ref{fig:goodrms}e. This model gave a fit to both the high and low flux power spectra with good $\chi^{2}$ statistics ($\chi^{2}$ = 35 and 47 respectively with 38 dof). The models used were broadly similar for both of the spectra. The main difference was in the first broad Lorentzian at high frequencies which was wider in the high flux spectrum, ($Q =$ 1.9 compared to 1.0 in the low flux spectrum). Other than this the only difference between the two models was in the normalisation of the components. The drop in the rms for the power law component between the high and low flux power spectra was 21$\%$, in contrast the rms of both the combined Lorentzians for the high frequency peaked component and the Lorentzian for the lowest frequency component dropped by 67$\%$. This suggests the existence of two separate components combining to produce the observed light curve, one forming the peaked noise and another the underlying broad-band component. Both of these components show a positive linear rms-flux relation, but have different flux-axis intercepts. This might help explain the well constrained negative intercepts observed in the rms-flux relations in some power spectra from this state (see Section \ref{sec:statedep}).

The other 5 observations which did not meet the criteria for inclusion (i.e. did not display a consistent flux dependence over the whole frequency range) were generally found to contain components shifting in frequency over the course of the observation, sometimes outside the 1-10 Hz range. These shifts, which have in these cases still produced strong, positively correlated, linear rms-flux relations, demonstrate the importance of careful analysis of the power spectra when measuring the rms-flux relation. Small differences in the shape of the PSD, whilst not always drawing it away from linearity, may artificially alter the gradient and thus the intercept of the observed rms-flux relation.


\section{Discussion}

\subsection{Summary}

\begin{figure*}
\begin{center}

	\includegraphics[width=14.0cm, angle=90]{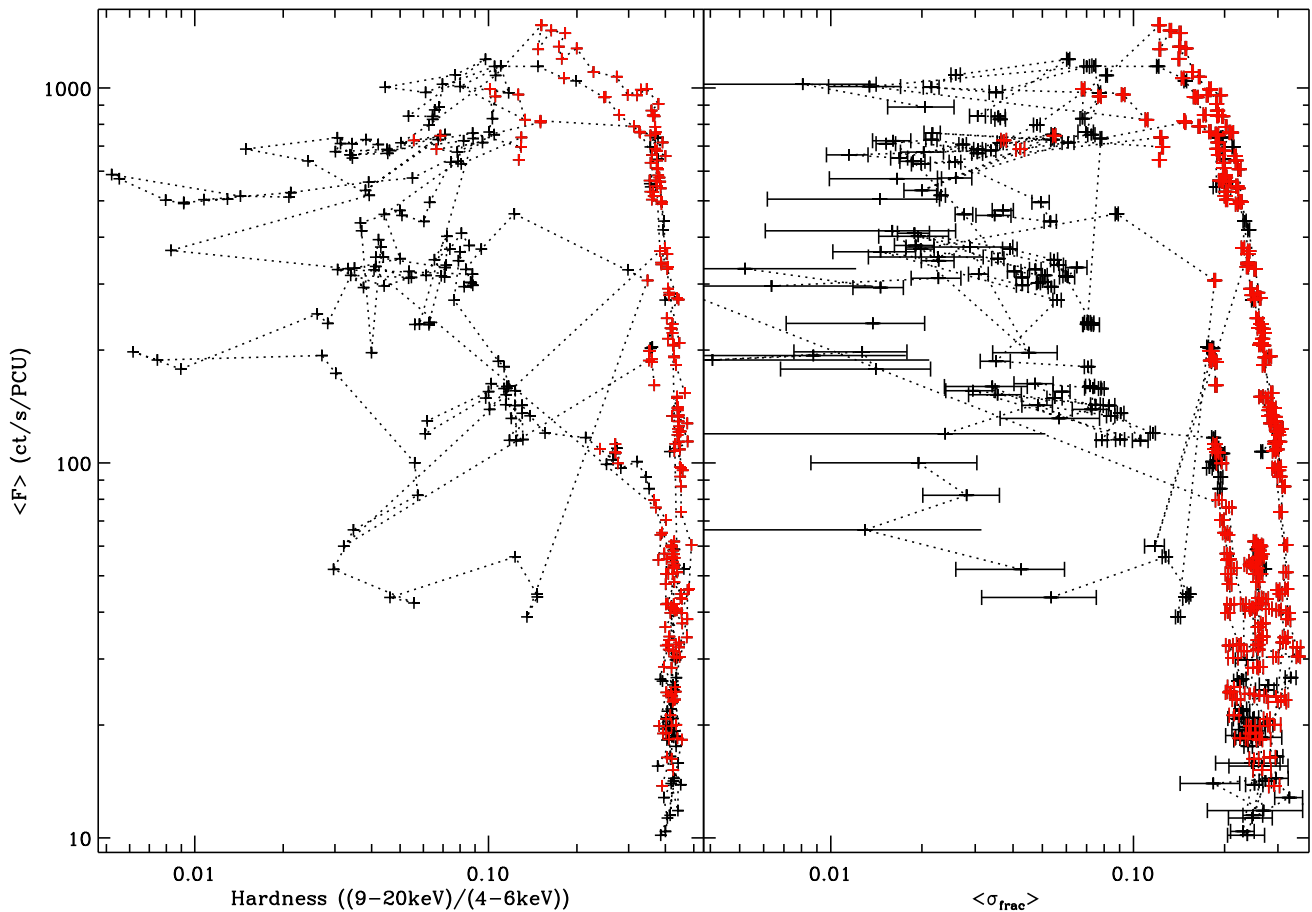}

\end{center}
\caption[]{HID (left) and RID (right) for all observations of GX 339-4. The ``good" observations are shown with red symbols, and black symbols indicate the data (mostly soft states) with rms below our threshold for rms-flux analysis.}

\label{fig:339all}
\end{figure*}

A large sample of observations from nine transient black hole X-ray binaries have been tested for a linear rms-flux relation. Analysis of the sample revealed the following results:

\begin{itemize}
	\item In every observation with sufficient flux and intrinsic variability to measure the rms, and a stationary PSD, a positive linear rms-flux correlation was observed in the 1-10 Hz range.
	\item All sources and states show long-timescale rms-flux relations (for observations with little alteration to power spectral shape within the 1-10 Hz band), similar to that already seen in Cygnus X-1. These bear close relation to those observed in Cygnus X-1 \citep{Gleissner04} and show remarkable similarities both in repeating outbursts of the same source, and in the behaviour of different sources \citep[i.e.][]{Munoz10}.
	\item There appears to be a link between $k$, $C_x$ and source state. A wide range is observed in the flux intercept ($C_x$). Hard states have x-axis flux intercepts close to zero, strong positive intercepts are observed in the HIMS. Strong negative intercept seem primarily isolated to the soft states and it is for observations in these states that the greatest range in $C_{x}$ is observed.
	\item There is little evidence for strong flux dependence in the PSD shapes apart from in some observations of an anomalous soft state of GRO J1655-40.
\end{itemize} 

\subsection{The ubiquity of the rms-flux relation}

The apparent ubiquity of the rms-flux relation in the broad-band noise within observations of these black holes implies that the coupling of variability over all timescales is common to the accretion process in all states. 

 Models similar to that explored in \cite{Lyubarskii97} where perturbations in the accretion rate propagate inwards through the accretion disc could provide an explanation for this behaviour. Variations on long timescales originating at larger radii in the accretion disc propagate inwards, coupling with faster perturbations at smaller radii, therefore creating a continuous relation between the mean accretion rate and amplitude of variations on all time scales, variations reaching inner regions modulate the X-rays. \cite{Uttley11} and \cite{Revnivtsev11} have studied observational evidence for the propagation of variability through the accretion disc. \cite{Uttley11} use data from GX 339-4 taken with \xmm~ which reaches softer energies than observations with \xte, in order to test for time lags between different energy bands. They find clear evidence that soft disc black body emission leads hard emission over long timescales ($>$ 1 s). The length of the observed lags suggest that they are generated by the inward propagation of accretion fluctuations through the X-ray emitting region. \cite{Revnivtsev11} observe some optical emission leading the X-rays in simultaneous observations of three intermediate polars, the lead-time ($\sim$ 7 s) agrees closely with that expected for the travel time of captured material onto the white dwarf. These results are part of an increasing array of evidence for the propagating fluctuations model, and suggest that a linear rms-flux relation should be expected in the broad band noise within observations of a whole range of accreting compact objects. 


\subsection{Interpreting state dependence of the intercepts} 

\begin{figure}
\begin{center}

	\includegraphics[width=6.7cm, angle=90]{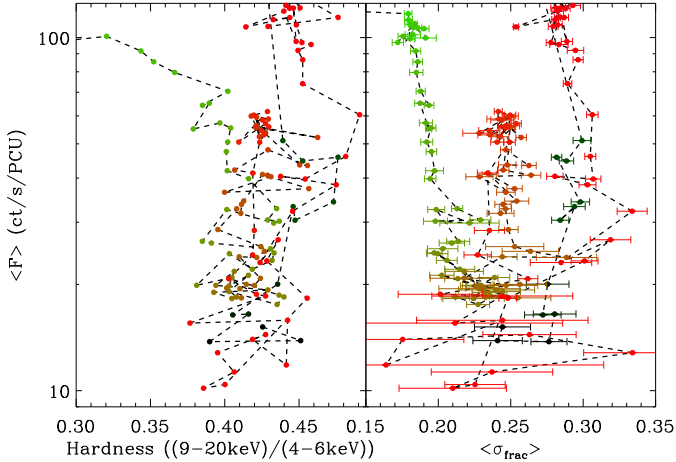}

\end{center}
\caption[]{HID (left) and RID (right) for GX 339-4. Observations shown are those from the 2007 outburst and beginning of that from 2010 which have count rates less that 100 ct s$^{-1}$. Points are colour coded according to MJD from green to red.}

\label{fig:339close}
\end{figure}

Plotting k and C$_{frac}$ against count rate reveals clear state dependence, when compared to both the HID and RID (see Figure \ref{fig:Intplts}). The RID shown here is similar to those first presented in \cite{Munoz10} and \cite{Motta11}, in this case however the x-axis is $\langle \hat{\sigma}_{frac} \rangle$, not simply $\langle \hat{\sigma} \rangle$. Figure \ref{fig:339all} shows all observations of GX 339-4 processed for this work with those used in the analysis marked in red. Lines of constant fractional intercept are diagonal in the plots presented by \cite{Munoz10} and vertical here.
The RID and the HID do not have identical
shapes (although the mapping between the two is very close).
Some features are not present in both plots --
for example, the RIDs for both GX 339-4 and XTE
J1550-564 show separate paths in the RID which are not
apparent in the HID. In the case of GX 339-4, as described in \cite{Munoz10}, the start of the
outburst and return to quiescence are along separate paths
in the RID (see Figure \ref{fig:339close}) although they appear to merge as the source fades. This behaviour is observed over repeated
outbursts but occurs at very similar values of the
hardness ratio and is therefore hidden in the HID. There
is also a third path visible in Figure \ref{fig:339close}, followed when the source flux rises
from low flux levels at the end of the 2007 outburst but then returns to quiescent levels. Similarly in XTE J1550-564 the return to
the hard state following an outburst where state changes
have occurred, is along a different route in the RID to
that observed for outbursts when the source remains in the
hard state. In the three hard outbursts XTE J1550-564 rises
and decays along very similar paths of fractional rms.
Again these tracks are much harder to discern on the HID.

Conversely, there are features visible in the HID which
are not as clear in the RID. For instance there are many
good observations of GRO J1655-500 taken when the source
was in an anomalous soft state; two separate types of
power spectra are visible whilst the source is in this
state (see e.g. \ref{fig:goodrms}e
and f). Those which display a typical soft state shape
(i.e. \ref{fig:goodrms}f) have a softer energy spectrum
than the observations which are more intermediate (i.e.
\ref{fig:goodrms}e), however the fractional rms in the
$1$--$10$ Hz range is very similar for both and this
difference
can only be observed in the HID. The use of both of these
diagrams in conjunction allows for a strong identification
of source state without the need to study details of the energy or
power spectra individually.
 
The well constrained positive and negative flux intercepts found for the rms-flux relation have been explained in previous work as secondary components within the light curve, \citep{Uttley01, Gleissner04}. Both \cite{Gleissner04} and \cite{Vaughan11} find evidence for energy dependence in the offset. \cite{Vaughan11} show that in observations of the AGN NGC 4051 the spectrum of the flux offset closely follows the soft component in the energy spectra, disappearing entirely at hard energies ($>$ 2keV). This could suggest that the strong positive intercepts observed in HIMS within the sample are due to a soft non-varying component contributing with increasing strength as the source transitions into a softer state. However this does not explain the wide range of behaviour visible in the anomalous soft states. Figure \ref{fig:Intplts}a shows values between -0.98 $<$ C$_{frac}$ $<$ 0.5, although the gradients remain low (k $<~0.1$), any additional components would have to be varying in strength and power quite considerably to create this range. The lack of flux dependence within the PSD shapes for most observations within the sample is also important when considering the origin of any additional components. It suggests that the power spectra of the components must have similar shapes. The outlying cases (i.e Figure \ref{fig:goodrms}e) may be caused by either the presence of two components contributing on different timescales, or one additional component contributing to a small frequency range and reacting differently to a change in the flux. Loosely classifying the component's frequency ranges into 0.1-1.0 Hz and 1.0-20.0 Hz for the case in Figure \ref{fig:goodrms}e the observed intercepts are very different. In the former range the fractional intercept is C$_{frac_{0.1-1.0}}$ = -6.61$\pm$1.4, in the latter C$_{frac_{1.0-20.0}}$ = 0.1$\pm$0.01 in both cases the likelihood of the intercept actually being zero is excluded at the 99$\%$ level. The different frequency ranges observed may also be indicative of the components originating from different locations, as it suggests that they dominate over different timescales. This wildly differing behaviour demonstrates that additional components within the lightcurve are likely to be present in at least some cases.

This work demonstrates that the rms-flux relation is a ubiquitous property of the broad band noise in light curves from accreting black holes and that the behaviour of this relation is clearly dependent on the state of the source. Further work on other compact objects, i.e. neutron stars and cataclysmic variable stars, AGN and ULXs, could reveal that this is in fact a ubiquitous property of the luminous accretion flow in all compact objects. 

\section*{acknowledgments}

We would like to thank the reviewer for their helpful comments. LMH acknowledges support from an STFC studentship. This research has made use of data obtained from the High Energy Astrophysics Science Archive Research Center (HEASARC), provided by NASA's Goddard Space Flight Center. 

\bibliographystyle{mn2e}
\bibliography{rmsbh_chapconv.bib}

\bsp


\appendix

\section{Error calculation for rms-flux points}
\label{app:errors}
Here we derive the formula used to estimate the standard deviation of
the estimates of the rms used in the rms-flux analysis. We have calculated them in a more general manner to that of \citep{Gleissner04}. Each time
series $x_t$, sampled at a rate $\Delta t$, is broken into segments of length $N_{\rm seg}$ and for
each segment a periodogram is computed, and the mean count rate
estimated, $\langle x^i \rangle$. The periodogram values are
calculated at Fourier frequencies $f_j = j / N_{\rm seg} \Delta t$
(for $j=1,2,\ldots,N_{\rm seg}/2$),
and the periodogram value at the $j$th frequency is $y_j^i$ for segment $i$.
The frequency resolution of such a periodogram is defined by the
length of the segments: $\Delta f = 1/N_{\rm seg} \Delta t$. 

The power spectrum at a given flux level is estimated by arranging 
the periodograms in order of the corresponding average
count rate $\langle x^i \rangle$, and averaging them in groups of $M$
periodograms.
For each count rate bin we calculate the average periodogram
\begin{equation}
  \langle y_j \rangle = \frac{1}{M} \sum_{i=1}^{M} y_j^i.
\end{equation}
It is well known that, in the absence of sampling distortions, 
the periodogram ordinates are distributed about the
true spectrum following a chi square distribution \citep[see][]{Jenkins69, Groth75, Priestley81, Leahy83, vanderKlis89, Press92, Percival93,
  Timmer95, Bloomfield00, Chatfield03}. Specifically, $y_j^i \sim P_j
\chi_2^2 /2$ where $P_j = P(f_j)$ is the true power density at
frequency $f_j$ and $\chi_2^2$ is a random variable with a chi square
distribution (with two degrees of freedom). 

The expectation and variance
of the chi square random variable are $\E [\chi_2^2 ] = 2$ and $\V [
\chi_2^2 ] = 4$. Using the rules for manipulating expectations and
variances \citep[e.g.][p23--24]{Eadie71} we have $\E[ y_j^i] = P_j$
and $\V[ y_j^i] = P_j^2$ and so
\begin{equation}
  \E[ \langle y_j \rangle ] = \frac{1}{M} \sum_{i=1}^{M} \E[y_j^i] = P_j,
\end{equation}
\begin{equation}
  \V[ \langle y_j \rangle ] = \frac{1}{M^2} \sum_{i=1}^{M} \V[y_j^i] = \frac{P_j^2}{M},
\end{equation}
where we have assumed the process is stationary, such that
$\E[ y_j^i] = P_j$ is the same for all segments $i$ (within the given
count rate bin), and the
individual $y_j^i$ values are distributed independently for all $i$. The above results
show that the averaged periodogram is an unbiased estimator of the true
spectrum $P(f_j)$ and its fractional ``standard error'' is $M^{-1/2}$.

An estimate of the variability power, $S$, contained within a specific
frequency range, say $f_{J}$ to $f_J + (W-1) \Delta f$, is given by 
summing the powers in the averaged periodogram $\langle y_j \rangle$ over the
appropriate frequencies:
\begin{equation}
  S = \sum_{j=J}^{J+W-1} \langle y_j \rangle \Delta f.
\end{equation}
Given the above information it is straightforward to calculate the
expectation and variance of the power estimate $S$:
\begin{equation}
\label{eqn:expS}
  \E[ S ] =  \Delta f \sum_{j} \E[ \langle y_j \rangle ] = \Delta f \sum_{j} P_j,
\end{equation}
\begin{equation}
\label{eqn:varS}
  \V[ S ] = \Delta f^2 \sum_{j} \V[ \langle y_j \rangle] = \frac{\Delta f^2}{M} \sum_{j} P_j^2,
\end{equation}
where we have assumed that the periodogram estimates, and their
averages $\bar{y}_j$, are independently
distributed for all $j$, which is asymptotically true (as $N_{\rm seg}
\rightarrow \infty$). These results show $S$ is an unbiased and
consistent estimator of the integrated power spectrum over the required
frequency range.

The estimated power $S$ includes a contribution from the Poisson noise
which is expected to contributed a flat spectrum with density
$P_N$. Over the frequency interval $W \Delta f$ it will therefore have a
total power of $P_N W \Delta f$. We may then calculate
the source (i.e. noise-subtracted) rms amplitude as
\begin{equation}
  \hat{\sigma} = (S - P_N W \Delta f)^{1/2},
\end{equation}
(assuming the noise level $P_N$ is
well determined so we can neglect any uncertainty in this).
Applying the usual rules for transformation of variances we find the
standard deviation of the estimator is
\begin{equation}
  {\rm sd}[ \hat{\sigma} ] = \frac{1}{2 \hat{\sigma}}
  \left( \frac{\Delta f^2}{M} \sum_{j} P_j^2 \right)^{1/2}
\end{equation}
Of course, we do not know the values of $P_j$, only their estimates
$\bar{y}_j$, so we obtain a reasonable estimate of the uncertainty on
the rms with the formula
\begin{equation}
  {\rm sd}[ \hat{\sigma} ] \approx \frac{\Delta f}{2 \hat{\sigma} M^{1/2}}
  \left( \sum_{j} \langle y_j^2 \rangle \right)^{1/2}
\end{equation}
This is essentially just the ``propagation'' (by adding in
quadrature) of the standard deviations of
each periodogram value, which are themselves $\sim y_j^i$, through the 
averaging over $M$ segments and summation over $W$ frequencies (each
of which is independently distributed).

This is different from the prescription used by
\cite{Gleissner04}. Their formula is only valid when the power is the
same in all frequency bins, i.e. locally white noise. When this is not the
case, e.g. when most of the power comes from a few dominant
frequencies (strong power laws or peaked QPOs), both the power estimate and the variance on the power
estimate are dominated by the frequencies that contribute the most power. 
One way to see this is to consider adding more frequencies that
contribute negligible power to the sum.
Using the \cite{Gleissner04} formula the relative error on the total power 
is $(MW)^{-1/2}$ (from their equation 2), whereas the above equations
predict this as $(\sum P_j^2/M)^{1/2} / \sum P_j$
(combining equations \ref{eqn:expS} and \ref{eqn:varS}). 
The former decreases as $W$ increases even if the additional
frequencies contribute no power, whereas the latter formulation leaves
the relative error unchanged, as expected if the additional data
contribute nothing. 
In general the \cite{Gleissner04} formula will underestimate the
errors, i.e. produce confidence intervals that are
smaller than the expected scatter in the rms data.

\label{lastpage}

\end{document}